\documentclass[notitlepage]{revtex4-1}%{iopart}
\pdfoutput=1
\usepackage[english]{babel}
\usepackage[pdftex]{graphicx}
\usepackage{multirow}
\usepackage{dcolumn}
\usepackage{bm}
\usepackage{mathbbol}
\usepackage{amsmath}
\usepackage{amsthm}
\usepackage{times}

\begin{document}

\title{Analytical description of the survival probability of coherent states in regular regimes}
%\title{Survival probability of coherent states in regular regimes}
\author{Sergio Lerma-Hern\'andez$^{1,2}$, Jorge Ch\'avez-Carlos$^2$, Miguel A. Bastarrachea-Magnani$^3$, Lea F. Santos$^4$,  Jorge G. Hirsch$^2$}
\affiliation{%\address{
$^1$ Facultad  de F\'\i sica, Universidad Veracruzana, Circuito Aguirre Beltr\'an s/n, C.P. 91000  Xalapa, Mexico\\%}
%\address{
$^2$ Instituto de Ciencias Nucleares, Universidad Nacional Aut\'onoma de M\'exico, Apdo. Postal 70-543, C.P. 04510  Cd. Mx., Mexico\\%}
%\address{$^3$ 
Physikalisches Institut, Albert-Ludwigs-Universitat Freiburg, Hermann-Herder-Str. 3, Freiburg, Germany, D-79104\\%}
%\address{
$^4$ Department of Physics, Yeshiva University, New York, New York 10016, USA}

%%%%%%%%%%%%%%%%%%%%%%%%%%%%%%%%%%%%%%%%
\begin{abstract}
Using coherent states as initial states, we investigate the quantum dynamics of the Lipkin-Meshkov-Glick (LMG) and Dicke models in the semi-classical limit. They are representative models of bounded systems with one- and two-degrees of freedom, respectively. The first model is integrable, while  the second one has both regular and chaotic regimes. Our analysis is based on the survival probability. Within the regular regime, the energy distribution of the initial coherent states consists of quasi-harmonic sub-sequences of energies with Gaussian weights. This allows for the derivation of analytical expressions that accurately describe the entire evolution of the survival probability,  from $t=0$ to the saturation of the dynamics. The evolution shows decaying oscillations with a rate that depends on the anharmonicity of the spectrum and, in the case of the Dicke model, on  interference terms coming from the simultaneous excitation of its two-degrees of freedom. As we move away from the regular regime, the  complexity of the survival probability  is shown to be closely connected with  the properties of the corresponding classical phase space. Our approach has broad applicability, since its  central assumptions are not particular of the studied models. 
%Since the central assumptions of our approach are not   particular for the studied models, it has broad applicability.   
\end{abstract}

\maketitle

%%%%%%%%%%%%%%%% INTRODUCTION %%%%%%%%%%%%%%%
\section{Introduction}

Highly controllable experiments with cold atoms~\cite{Chu2002,BernienARXIV}, ion traps~\cite{Blatt2012,Richerme2014}, and nuclear magnetic resonance (NMR) platforms~\cite{WeiARXIV}, where coherent evolution can be investigated for long times, are in part responsible for the renewed interest in non-equilibrium quantum dynamics. %Several paradigmatic models of many-body quantum physics have become experimentally accessible. The
Alongside with several paradigmatic models of many-body quantum physics, simple but rich ones like the
 Lipkin-Meshkov-Glick (LMG)~\cite{Lipkin1965a,Lipkin1965b,Lipkin1965c} 
and the Dicke~\cite{Dicke54,Hepp73,Wang1973} 
models have become experimentally accessible.
%, for instance, 
The latter were realized with Bose-Einstein condensates in
%Ref.~
~\cite{Milburn1997,Steel1998,Zibold2010} and 
%Refs.~
~\cite{Baumann2010,Baumann2011}, respectively. 

To better understand and control many-body quantum systems out of equilibrium, in addition to experimental and numerical studies, one 
%needs 
can exploit the advantages of 
analytical results %that 
%can 
%help us
 to identify and  explain the causes of different behaviors at different time scales. However, analytical results are challenging in systems that approach chaotic regimes.

The present work focuses on the analytical description of the equilibration process of the LMG and Dicke models. They are representative models of bounded systems with one- and two-degrees of freedom, respectively (the number of degrees of freedom defined through the classical limit). The LMG model is integrable, while the Dicke model presents both regular and chaotic classical trajectories. Our analysis concentrates on the regular regime, which enables the derivation of analytical expressions that cover the entire dynamics of the two systems. In the case of the Dicke model, by gradually moving the initial state away from the regular regime, we are able to identify the source of the increased complexity of the dynamics.

The quantity that we select for our studies is the survival probability ($SP$), that is the probability of finding the initial state later in time.   
%SER 
$SP$  is a simple dynamical quantity that encodes  the structure of the energy components of the initial state, making it a valuable tool to detect and study  critical phenomena in the energy spectrum, such as quantum phase transitions (QPT)~\cite{Campbell16}, excited-state quantum phase transitions (ESQPT)~\cite{Wang17,SantosBernal2015,Santos2016, Fernandez2011} and dynamical phase transitions~\cite{DQPT,DQPT2};  correlations in the energy spectrum that  distinguish between regular and chaotic systems~\cite{Alhassid92,Torres2017b}; decay of unstable systems~\cite{Fonda78}; metal-insulator transition~\cite{Ketz92}; and  quantum speed limit~\cite{Khalfin58}, among other subjects.

The survival probability (also known as return probability) and the Loschmidt echo~\cite{Peres1984,Gorin2006} are particular cases of the fidelity between two pure states. While the survival probability measures the overlap between the initial state and its evolved counterpart, the Loschmidt echo evaluates the overlap between the initial state evolved under two different Hamiltonians. %Equivalently, as routinely done in NMR experiments, the Loschmidt echo measures the overlap between an initial state and its evolved counterpart which faces both an evolution forward under a certain Hamiltonian and then backward according to a different Hamiltonian, therefore the name ``echo''.
 In the scenario of small perturbations, where the two Hamiltonians are only slightly different, analytical expressions for the Loschmidt echo have been obtained~\cite{Prosen2002,Prosen2003}. We stress that our focus is on the survival probability and on very strong perturbations that take the system far from equilibrium.

As discussed in previous works, at  short times the $SP$ shows a universal quadratic decay with rate  determined by the energy variance of the initial state. Its subsequent decay is controlled by the shape of the energy distribution, Gaussian and exponential behaviors being common for strong perturbations~\cite{*Torres2014PRA,*Torres2014NJP,*Torres2014PRE,Torres2014PRAb, Flamabum2001,Izrailev2006}. At later times, the $SP$ behavior is rather complex, depending strongly  on the details of the energy components probed by the initial state~\cite{%Torres2015,
*Tavora2016,*Tavora2017, Alhassid92,Torres2017}. For finite-size systems the $SP$ eventually  saturates to   its infinite-time average at the equilibration time, showing fluctuations whose temporal dispersion is of the order of the saturation value.

In the present paper, the $SP$ is used to describe in detail the different temporal scales in the equilibration process of regular quantum systems with few-degrees of freedom and with a well defined classical limit. To gain insights from the classical dynamics, we use coherent states~\cite{KlauderBook, Gilmore90} as initial states and consider a small effective Planck constant \cite{Altland2012NJP,Altland2012PRL,Bakemeier13}.

Our approach finds inspiration in the study of  multilevel quantum beats  in~\cite{Leichtle1996}.  We identify the relevant properties of the initial state and the energy spectrum responsible for the dynamic behaviors observed at different times. This analysis enables us to provide  a precise definition of the equilibration time.  We stress that our approach  
can be extended to other similar models where the spectrum has a regular part.  It was indeed recently employed  in~\cite{Kloc18} for the analysis of the quenched dynamics of the integrable two-degrees of freedom Tavis-Cummings model and in~\cite{HUMBERTO} for a one-dimensional quartic double-well potential in the semi-classical limit.  

In the LMG model, the analytical expression that we obtain for the $SP(t)$  is a sum of products of cosine and Gaussian functions. It depends only on three parameters that can be  estimated analytically and semi-classically. The decay rate of the oscillations of the $SP(t)$ is proportional to the anharmonicity of the spectrum  probed by the initial state. 
In the Dicke model, since  the regular part of the energy spectrum is organized in %sets of eigenstates that form
 invariant subspaces associated with the quantum numbers of approximate integrals of motion~\cite{Relano16,BastaJPA17}%. As a consequence
, instead of a single sum, %, as in the case of the LMG model,
 the analytical expression for the $SP(t)$ consists of different sums and interferences between them. The number of sums grows as the energy and parameters of the initial coherent state approach chaotic classical regions. This causes the decay time of the oscillations to decrease significantly.  For both models, the analytical results are compared with numerics, showing  remarkable agreement.

The paper is organized as follows. Section~\ref{Sec:Hams} offers a brief presentation of the Hamiltonians and coherent states employed. In section~\ref{Sec:LMG}, we derive an analytical expression for the survival probability evolving under the LMG model. In section~\ref{Sec:DMSP}, the analytical expression for the survival probability obtained with the LMG model is generalized to describe the regular regime of the Dicke model. Conclusions are given in section~\ref{Sec:Concl}. In addition, several appendices provide details of the derivations.

%%%%%%%%%%%%%%%% HAMILTONIAN% %%%%%%%%%%%%%%%
\section{Hamiltonians, Initial States, and Survival Probability}
\label{Sec:Hams}
The LMG and Dicke models were proposed with the common motivation of providing schematic models capable of capturing essential phenomena of many-body quantum physics: the transition between the spherical and deformed phase of nuclei, in the case of the LMG model, and the interaction between radiation and matter for the Dicke model. Both describe the interaction of $N$ two-level systems, mutually interacting in the case of the LMG model, while in the Dicke model they are
% interact with 
coupled to a single bosonic mode of frequency $\omega$. 

The Hamiltonian that describes the LMG model is given by
\begin{equation}
\hat{H}_{LMG}=\hat{J}_z+ \frac{\gamma_x}{2J-1} \hat{J}_x^2+ \frac{\gamma_y}{2J-1} \hat{J}_y^2,
\label{LMG}
\end{equation}
where $\hbar=1$.
For the Dicke model,
\begin{equation}
\hat{H}_{D}=\omega \hat{a}^{\dagger}\hat{a}+\omega_{0}\hat{J}_{z}+ \gamma\sqrt{\frac{2}{J}}\hat{J}_{x}\left(\hat{a}+\hat{a}^{\dagger}\right).
\label{Dicke}
\end{equation}
The pseudo-spin operators $\hat{J}_i$  satisfy the usual %$SU(2)$ 
$su(2)$ algebra,  with invariant subspaces labelled by the pseudospin quantum number $J$. The bosonic annihilation (creation) operator is $\hat{a}$ $(\hat{a}^\dagger)$, $\gamma_{x,y}$ is the coupling strength between the two-level systems and $\gamma$ is the coupling strength between the field and the two-level systems.

Both models present a second-order ground state  QPT at critical values of their coupling constants.  For the LMG model~\cite{Castanos2005}, $\gamma_x^{cr}=-1$ for $\gamma_y\geq -1$ or $\gamma_y^{cr}=-1$ for $\gamma_x\geq -1$, and  for the Dicke model~\cite{Emary2003PRL,Emary2003},  $\gamma^{cr}=\sqrt{\omega\omega_o}/2$. The critical values separate  a normal phase (which includes the zero coupling cases) from a deformed (LMG) or superradiant (Dicke) phase. 
The LMG and Dicke Hamiltonians have a discrete parity symmetry, which separates the Hilbert
space in two invariant subspaces.

\subsection{Initial States and classical Hamiltonians}

Bloch and Glauber coherent states  ($z,\alpha \in \mathbb{C}$)~\cite{KlauderBook}
$$
|z\rangle=\frac{1}{\left(1+\left|z\right|^{2}\right)^{J}} e^{z \hat{J}_+}|J, -J\rangle, {\hbox{ \ \ \ and \ \ \ \ }} |\alpha\rangle=e^{-|\alpha|^2/2}e^{\alpha \hat{a}^\dagger}|0\rangle,
$$
are used as initial states for the  LMG, $|\Psi(0)\rangle=|z_0\rangle$,  and the Dicke, $|\Psi(0)\rangle= |z_0\rangle\otimes |\alpha_0\rangle$, models. Likewise they are used to define classical corresponding Hamiltonians \cite{Gilmore90,Ribeiro2006}: $h_{LMG}=\langle z|H_{LMG}|z\rangle/J$ and $h_{D}=\langle z|\otimes \langle \alpha | H_D | \alpha\rangle\otimes |z\rangle/J$. 

This choice of initial states is natural when one wants to make a clear connection between the results of the quantum dynamics with the properties of the classical phase space. Indeed, the canonical classical  variables $(\phi,j_z)$ and $(q,p)$ are  given in terms of the coherent state  parameters
$$
z=\sqrt{\frac{1+j_z}{1-j_z}}e^{-i \phi}
%z=\sqrt{\frac{1+j_z/J}{1-j_z/J}}e^{-i \phi}
\hbox{ \ \ \ and \ \ \ }
\alpha=\sqrt{\frac{J}{2}}(q+ip).
$$
The classical limit is obtained by considering $J\rightarrow \infty$~\cite{Larson17}, the effective Planck constant being $\hbar_{eff}=1/J$. We choose initial states  in regular regions of the corresponding classical phase space, in the deformed (LMG) or superradiant (Dicke) phases. We also use      positive-parity  projected~\cite{CastanosProceed} initial states, although this choice is not crucial. %for our approach applies.

In addition to regular dynamics, the other important criterion for our analysis  is that  the initial coherent states have marginal or null components of energy levels from critical energy regions, that is ground-state and ESQPT~\cite{Caprio2008,Ribeiro2008,Bastarrachea14PRA-1} energies. The latter critical phenomenon is common in few-degrees of freedom models~\cite{Stransky2014,Stransky2015}.
Studies of the effects of an ESQPT in the temporal evolution of the LMG and Dicke models  include~\cite{SantosBernal2015,Santos2016,Bernal2017,Engelhardt2015} and~\cite{Fernandez2011}, respectively.  We leave out from this contribution the analysis of these critical cases. We emphasize that the results presented in this work are general for coherent initial states away from critical points. In addition, they are valid not only to the LMG and Dicke models, but also to other models with a regular part of the spectrum, such as those in \cite{Kloc18,HUMBERTO}.

% For both models,  the selected energy regions present pair-wise degeneracies     of  positive and negative parity states \cite{Puebla2013}. To avoid tunnelling issues related to these degeneracies and simplify the discussion, we consider   coherent states projected  in the positive parity sector~\cite{CastanosProceed}.

%%%%%%%%%%%%%%%%%%%%%%%%%%%%%%%%%%%%%%%%
\subsection{Numerical method}
The numerical results for the dynamics are obtained by exactly diagonalizing the Hamiltonians and decomposing the initial state in the positive parity energy eigenstates $|E_k\rangle$, so that the evolved state is
\begin{equation}
|\Psi(t)\rangle = \sum_k c_k e^{-i E_k t} |E_k\rangle.
\end{equation}
Above, $E_k$ are the eigenvalues of the Hamiltonian and
%\begin{equation}
$c_k = \langle E_k | \Psi(0)\rangle$ 
%\end{equation}
are the numerically evaluated  overlaps between the initial state and the positive parity eigenstates.

For the LMG model, where the size of the Hilbert space is finite, we can consider  relatively large pseudospin values and thus explore, without much computational effort, the convergence to the classical limit. We select $J=2000$, $\gamma_x=-3$, and $\gamma_y=-5$.

For the Dicke model, the unbounded number of bosons makes the Hilbert space infinite.  In order to diagonalize its Hamiltonian, a truncation in the  number of bosonic excitations is introduced. The cut off has to be large enough to guarantee convergence of the low energy results that we are interested in. We use the   basis described in~\cite{Chen2008,Liu2009,Bastarrachea2014PSa,Bastarrachea2014PSb} to diagonalize the Hamiltonian. This basis is particularly efficient to obtain, in the superradiant phase, rapid convergence of a large portion of the low-energy spectrum as a function of the cut off. However, the values of $J$ computationally affordable are much smaller than in the LMG model.
 We use  $J=120$ and  consider a resonant case $\omega=\omega_0=1$ with the coupling strength   
$\gamma=2\gamma_c=\sqrt{\omega\omega_0}=1$. The technical details to calculate  the energy components of the initial coherent states can be found in Appendix C of~\cite{Bastarrachea2016a}.

%%%%%%%%%%%%%%%%%%%%%%%%%%%%%%%%%%%%%%%%
\subsection{Survival Probability}

%Expanding the initial state $|\Psi(0) \rangle $ in the Hamiltonian  eigenstate basis $|E_k \rangle $ with eigenergies $E_k$,
%$|\Psi(0) \rangle = \sum\limits_{k} c_k |E_k \rangle$,
 The survival probablity $SP(t)=\left| \langle \Psi(0) |\Psi(t) \rangle \right|^2$ can be written as
\begin{equation}
SP(t) = \displaystyle \sum_{p=1} SP_p(t) + IPR
\label{SPdef}
\end{equation}
where
\begin{equation}
SP_p(t) \equiv \displaystyle\sum_{k}  2|c_{k+p}|^2|c_k|^2  \cos(\omega_k^{(p)}t)  ,
\label{SPp}
\end{equation}
\begin{equation}
\omega_k^{(p)} \equiv E_{k+p}-E_{k} ,
\label{Eq:omegaKp}
\end{equation}
and the index $p$ designates the distance between the eigenenergies. The sum for $p=1$ considers only nearest neighboring eigenvalues, the sum for $p=2$ only the second neighbors, and so on.

The inverse participation ratio, $IPR =\sum_{k}|c_k|^4$, is the infinite-time average of the survival probability. It measures the level of delocalization of the initial state in the energy eigenbasis. The dispersion of the temporal fluctuations of $SP(t)$ around $IPR$ is also  
%$\sim IPR$
of the order of the $IPR$~\cite{TorresKollmar2015}.

%%%%%%%%%%%%%%%% LMG MODEL %%%%%%%%%%%%%%%%%

\section{Survival probability in one-degree-of-freedom bounded systems}
\label{Sec:LMG}

The main result of this section is the analytical expression for the survival probability presented in (\ref{SPsuman}) of section~\ref{Subsec:AnalytExp}. Also important is the excellent agreement with the numerics shown in section~\ref{Subsec:CompNum}. The analysis presented for the LMG model can be extended to other Hamiltonians with one-degree of freedom, provided they have a discrete spectrum (bounded systems) and the mean energy of the initial state is far enough from critical energies (ground-state and ESQPTs).

%The initial state that we choose to evolve with the LMG model is a coherent state with 
As a representative example, we choose the initial state whose coordinates in the classical phase space are
%$j_{zo}/j
$j_{zo}
=-\cos(\pi/3)$ and $\phi_o=\pi/2$.  It has mean energy $\bar{E}/J=\sum_{k} |c_k|^2 E_k/J =-2.376$ and energy distribution of width $\sigma/J= [\sum_{k} |c_k|^2 E_k^2 - (\bar{E})^2]^{1/2}/J
=0.02054$.
%which can be evaluated  numerically or   analytically [see (\ref{eq-app2:sigLMG}) in \ref{App2}%Section C in \cite{SupMat}]. 
The eigenstates of $\hat{H}_{LMG}$ that significantly contribute to the dynamics are
% therefore below the critical energy of the ESQPT, $E_{ESQPT}=J(\gamma_y+\gamma_y^{-1})/2=- 1.6667 J $ and 
in the low-energy region, but
far from  the ground-state  energy, $E_{GS}=J(\gamma_x+\gamma_x^{-1})/2=- 2.6 J$ and  the critical energy of the ESQPT, $E_{ESQPT}=J(\gamma_y+\gamma_y^{-1})/2=- 1.6667 J$~\cite{Ribeiro2008}. 
%%%%%%%%%%%%%%%%%%%%%%%%%%%%
\subsection{Components of the initial state and $IPR$} %$\pmb{IPR}$}

\begin{figure}
\begin{tabular}{cc}
(a) &(b)\\ 
\includegraphics[width=0.45\textwidth]{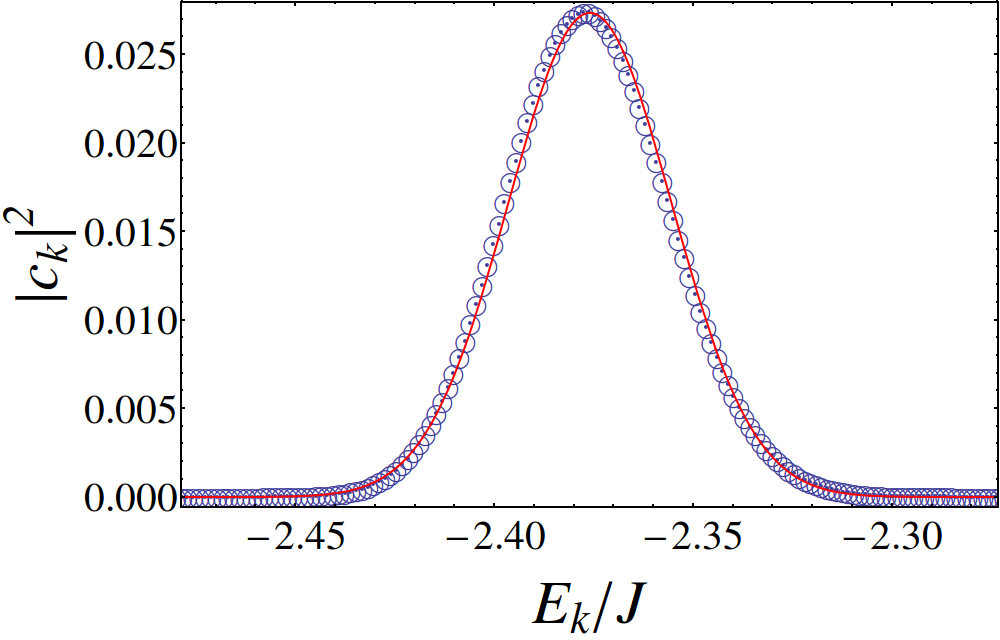} & \includegraphics[width=0.45\textwidth]{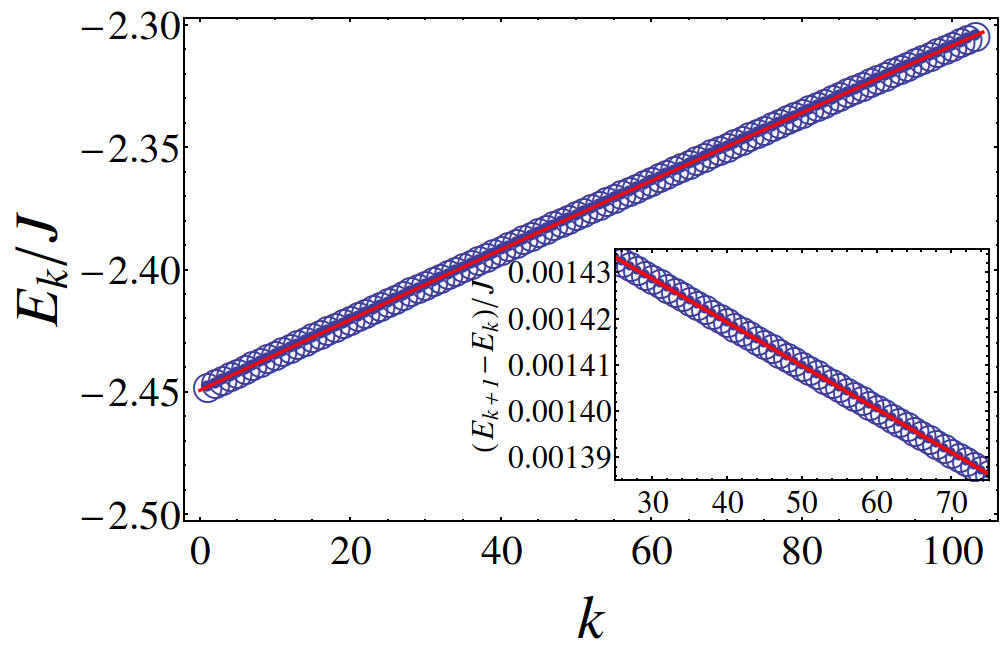}\\
 \multicolumn{2}{c}{(c)}\\
 \multicolumn{2}{c}{\includegraphics[width=0.6\textwidth]{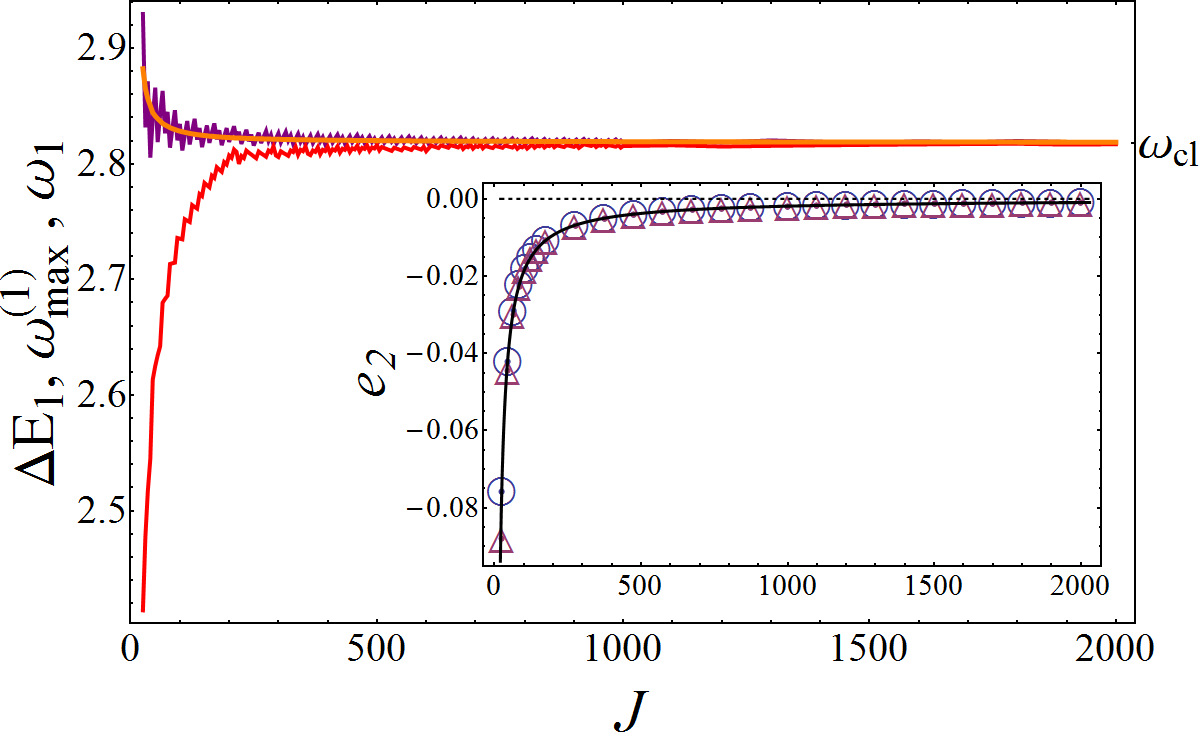}} 
\end{tabular}  
\caption{(a)  Absolute squared components (circles)  of the chosen coherent state in the energy eigenbasis of the LMG model %(see text) and the 
and its
Gaussian approximation using $\Delta E_1\approx \omega_1$ (solid line). 
(b) Eigenvalues (circles) in the interval $[\bar{E}-3.5\sigma,\bar{E}+3.5]$ plotted  against their ordering numbers in this region. The quadratic fit (\ref{principalspectrum}) is shown with a solid line. The inset displays the energy differences of consecutive eigenenergies and their fit.
(c) Mean value $\Delta E_1$ (red line increasing for small $J$)  of the differences of consecutive eigenergies in the same interval as the main panel in (b), $\omega_{max}^{(1)}=E_{kmax+1}-E_{kmax}$ (dark purple line fluctuating in its decay),  and $\omega_1$ evaluated from %Eq.~
(\ref{Eq:wp}) (light orange line). The asymptotic value of the three lines, which is the frequency of the classical model $\omega_{cl}=2.818$, is indicated on the right. The inset shows the anharmonicity $e_2$ as a function of $J$ calculated numerically from %Eq.~
(\ref{e2es}) (circles) and from the quadratic fit (triangles). The solid line is  the semi-classical approximation in %Eq.~
(\ref{e2sc}). Parameters:
 %$j_{z o}/J
 $j_{z o}
 =-\cos(\pi/3)$, $\phi_o=\pi/2$, $\bar{E}/J=-2.376$; in (a) and (b): $J=2000$.}
\label{fig:gaussLMG}
\end{figure}

In figure \ref{fig:gaussLMG} (a), we show the absolute squared components $|c_k|^2$  as a function of the eigenvalues of the LMG model. The components are very well approximated by a Gaussian function
\begin{equation}
|c_k|^2\approx g_{k}\equiv A e^{-\frac{(E_k-\bar{E})^2}{2\sigma^2}},
\label{components}
\end{equation}
as depicted in the figure with a solid line.
From the normalization condition, the amplitude $A$ can be shown to be
% (see \ref{App3} %Section D of  \cite{SupMat} for a detailed derivation)  
\begin{equation}
A=\frac{1}{\sqrt{2\pi}}\frac{\Delta E_1}{\sigma}, 
\label{amplitude}
\end{equation}
where 
\begin{equation}
\Delta E_1= \langle E_{k+1}-E_k \rangle
\end{equation} 
is the mean of the energy differences between consecutive energies of the states that contribute to the evolution of the coherent state. Concretely,  we consider  energy states in   the interval $[\bar{E}-3.5 \sigma, \bar{E}+3.5 \sigma]$, where lies $99.95\%$ of the norm of the initial state.

The infinite-time average of the survival probability is therefore given by
\begin{equation}
IPR=   \displaystyle\sum_k  |c_k|^4\approx A^2   \displaystyle\sum_k   e^{-\frac{(E_k-\bar{E})^2}{\sigma^2}}
\approx \dfrac{A^2}{\Delta E_1}   \displaystyle\int   e^{-\frac{(E-\bar{E})^2}{\sigma^2}}d E= \dfrac{1}{2\sqrt{\pi}}\dfrac{\Delta E_1}{\sigma}.
\label{IPR}
\end{equation}
As discussed in  \cite{Schliemann2015} (see also \ref{App2}), the
standard deviation of the energy distribution of coherent states is $\sigma\propto \sqrt{J}$. With this and from the fact that $\Delta E_1$ tends to a finite value in the limit $J \rightarrow \infty$ [see %Eq.
(\ref{limclassom}) below],  expression (\ref{IPR})  explains the results of %Ref.~ 
\cite{Bastarrachea2016a,BastarracheaPS2017}, where it was shown that the $IPR$ of coherent states in regular regions scales as $1/\sqrt{J}$ for large $J$. 

%%%%%
%%%%%

%%%%%%%%%%%%%%%%%%%%%%%%%%%%%%%%%%%%%%%%%%
\subsection{Frequencies and their  Distribution}

In search of an analytical expression for $SP(t)$, we now concentrate on the two key elements of $SP_p(t)$ in %Eq.~
(\ref{SPp}), namely the frequencies $\omega_k^{(p)}$ and their distribution given by the product $|c_{k+p}|^2|c_k|^2$, starting with the first ones.

\subsubsection{Frequencies $\omega_k^{(p)}$}
In figure~\ref{fig:gaussLMG} (b), the LMG eigenergies in the interval $[\bar{E}-3.5 \sigma, \bar{E}+3.5 \sigma]\sim [ -2.448 J,-2.304 J]$ are plotted with blue circles against their ordering numbers in this region. 
%This interval was selected, because the squared components of the energy eigenstates in it  account for around 99.95\% of the norm of the initial state. 
%Within this interval lies 99.95\% of the norm of the initial state. 
We show with a solid line that the data can be very well fitted with the semi-classical  expansion (see \ref{App6} for a detailed derivation)
\begin{equation}
%E_k^{(fit)}=e_o+e_1 k+e_2 k^2,
E_k = e_o+e_1 k+e_2 k^2,
\label{principalspectrum}
\end{equation}
where $k$ is an  integer number. This leads to  
\begin{equation}
%\omega_k^{(p)}=(E_{k+p}-E_{k})\approx (E_{k+p}^{(fit)}-E_{k}^{(fit)})= p(e_1 + p \,e_2)+ 2 \,e_2  \,p \,k.
\omega_k^{(p)}=(E_{k+p}-E_{k}) = p(e_1 + p \,e_2)+ 2 \,e_2  \,p \,k.
\label{omegap}
\end{equation}
%As discussed in %Ref.~\cite{Leichtle1996}, expression (\ref{principalspectrum}) can be viewed as a series in the variable $\hbar$  ($\hbar_{eff}=1/J$ here), where we have neglected the higher order terms. 
The anharmonicity $e_2$ measures the departure from a spectrum with equally spaced energies. 
%For the selected part of the LMG spectrum, this quadratic contribution is very small. In fact, the fitting curve in figure~\ref{fig:gaussLMG} (b) has 
It is very small,   $e_2=-0.00094$, when compared with $e_0=-4898.46$ and $e_1= 2.91$. The inset of figure~\ref{fig:gaussLMG} (b) shows the energy differences of consecutive eigenenergies (circles) and the result for
% $E^{(fit)}_{k+1}-E^{(fit)}_{k}
 $E_{k+1}-E_{k}
  =(e_1+e_2)+2 e_2 k$ (line), whose  slope is given by $e_2$. Despite small, $e_2$ has an important role in the decay of the survival probability, as will become clear later.
Equation~(\ref{principalspectrum}) is a valid assumption for  any coherent state in the semi-classical limit,  provided the energy interval defined by its mean energy and width does not include critical energies. 
%The exceptions are states very close to the ground state and in the vicinity of an ESQPT. 
%The divergence of the density of states associated with ESQPTs prevents approximating  the spectrum with a smooth sequence as the one in %Eq.
%(\ref{principalspectrum}). Studies of the effects of an ESQPT in the temporal evolution of the LMG 
%and Dicke models  include Refs.~\cite{SantosBernal2015,Santos2016,Bernal2017,Engelhardt2015} and \cite{Fernandez2011}, respectively.  We leave out from this contribution the analysis of these critical cases. 

%%%%%%%%%%%%%%%%%%%%%%%%%%%%%%%%%%%%%%%%
\subsubsection{Product $|c_{k+p}|^2|c_k|^2$}

Following %Eq.~
(\ref{components}), $|c_{k+p}|^2 |c_k|^2\approx g_{k+p} \, g_{k}$. \ref{App5} %Section E in \cite{SupMat} 
shows that this product can be very well approximated by a Gaussian distribution for frequencies $\omega_{k}^{(p)}$
\begin{equation}
g_{k+p} \, g_{k}
\approx  A_p  \exp\left[-\frac{\left(\omega_k^{(p)}-\omega_{p}\right)^2}{2\sigma_p^2}\right],
\label{product}
\end{equation} 
%Since $|e_2|<< |e_1|$, at leading order in $e_2$,
where the centroid  ($\omega_p$),  amplitude ($A_p$), and width ($\sigma_p$) % of the $p$-th distribution
 are given %simply 
 in terms of the values for~$p=1$ %(see \ref{App5}%Section E in \cite{SupMat}
%),
\begin{align}
\omega_{p} \approx p \,\omega_{1}, & \hspace{0.9 cm} \omega_{1} \approx\sqrt{e_1^2+4 e_2 (\bar{E}-e_0) } ,
\label{Eq:wp}
\\
\frac{A_p}{A^2}\approx \left( \frac{A_1}{A^2} \right)^{p^2}  , & \hspace{0.9 cm} \frac{A_1}{A^2} =  \exp\left(-\frac{\omega_{1}^2}{4\sigma^2} \right) ,
\label{Eq:Ap}
\\
\sigma_p \approx p \,\sigma_{1} , & \hspace{0.9 cm} \sigma_{1} = \sqrt{2}\, |e_2| \frac{\sigma}{\omega_{1}} .
\label{eq:aproxDis}
\end{align}
Therefore, the dominant frequency of the $p$-th component of the survival probability is approximately a harmonic frequency,  $\omega_1$ being the fundamental one.
The value $\omega_1$ is  where the product of Gaussians  $g_{k+1}g_k$ takes its maximal value. 
It is approximately  given by the pair of consecutive eigenenergies  
located around $\bar{E}$
\begin{equation}
\omega_{max}^{(1)} = E_{kmax+1}-E_{kmax},
\label{w1es}
\end{equation}
where the pair  $E_{kmax}$ and $E_{kmax+1}$ is defined through the condition $E_{kmax}\leq \bar{E} \leq E_{kmax+1}$.

To determine $A_1$, in addition to $\omega_1$, we also need $\Delta E_1$ through $A$ from (\ref{amplitude}). 
Since $\Delta E_1$ is the mean value of the differences of  consecutive energies in an interval around $\bar{E}$ and %since 
these differences vary linearly in this interval [cf. the inset of figure~\ref{fig:gaussLMG} (b)], we can approximate $\Delta E_1$ by the energy difference in the center of the interval, %that is $\omega_{max}^{(1)}$, 
\begin{equation}
\Delta E_1\approx\omega_{max}^{(1)}\approx \omega_1.
\label{iden}
\end{equation}
This  assumption is not exact, but  the three quantities converge, in the limit $J\to \infty$, to the classical frequency $\omega_{cl}$ 
\begin{equation}
\lim_{J\rightarrow \infty}  \Delta E_1 = \lim_{J\to\infty}\omega_{max}^{(1)}= \lim_{J\to\infty}\omega_1=\omega_{cl},
\label{limclassom}
\end{equation}
as shown in figure~\ref{fig:gaussLMG} (c) and  
%of the trajectory with initial conditions defined from the parameter $z_o$ of the initial coherent state% differences between  $\Delta E_1$, $\omega_{max}^{(1)}$ and  $\omega_1$ go to zero as we approach the classical limit, $J\to \infty$
%, as shown in figure~\ref{fig:gaussLMG} (c) and %In this limit, the anharmonicities become negligible, and a single frequency fully characterizes the classical trajectories.  
%The classical frequency is evaluated using  angle-action variables, as
 discussed in \ref{App6}.

It remains to find the width $\sigma_1$ in %Eq.~
(\ref{eq:aproxDis}), and for this we need $e_2$. The anharmonicity  
%can be obtained from the fitting of the spectrum, as discussed above. It can also be estimated, as we do here, 
is estimated
using assumption (\ref{principalspectrum}),
\begin{equation}
e_2=\frac{E_{kmax+1}+E_{kmax-1}}{2}-E_{kmax}.
\label{e2es}
\end{equation}
Small differences exist between $e_2$ estimated with the expression above and the anharmonicity obtained by fitting the spectrum with %Eq.~
(\ref{principalspectrum}), but %the discrepancy decreases and 
both values go to zero as $J$ increases and converge to the semi-classical (see \ref{App6}) expression   
\begin{equation}
 \lim_{J\rightarrow\infty} e_2= %\lim_{J\to \infty}-\frac{\omega_{cl}^3}{4\pi J}\frac{d^2 I}{d \epsilon^2}
 \frac{\omega_{cl}}{2 J}\frac{d \omega_{cl}}{d \epsilon}\equiv f_e/J \ \  \  \ \ \ \ {\hbox{(with \ \  $\epsilon=E/J$)}},
 \label{e2sc}
\end{equation}
as seen in the inset of figure~\ref{fig:gaussLMG} (c).
%%%%%%%%%%%%%%%%%%%%%%%%%%%%%%%%%%%%%%%%
\subsection{Analytical Expression}
\label{Subsec:AnalytExp} 
Putting the above results together in %Eq.~
(\ref{SPp}), we have
\begin{equation}
SP_p(t)\approx \frac{\omega_1^2 }{\pi \sigma^2}  \exp\left(-\frac{p^2  \omega_1^2}{4 \sigma^2} \right)
\sum_k   \exp\left[{-\frac{\left(\omega_k^{(p)}-p \omega_1\right)^2}{2 p^2 \sigma_1^2}} \right]\cos(\omega_k^{(p)}  t) .
\label{SPppp}
\end{equation}
Approximating the sum above by an integral (see \ref{App7} % Section F of \cite{SupMat}
 for details), we arrive at
%\begin{equation}
%SP_p(t)\approx \frac{\omega_1}{\sigma  \sqrt{\pi}}  \left[\exp \left(-\frac{ \omega_1^2}{4 \sigma^2} -  \frac{t^2}{t_D^2}\right)\right]^{p^2} \cos(p \omega_1  t),
%\end{equation}
\begin{equation}
SP_p(t)\approx \frac{\omega_1}{\sigma  \sqrt{\pi}}  \exp \left[-p^2\left(\frac{ \omega_1^2}{4 \sigma^2} +  \frac{t^2}{t_D^2}\right)\right] \cos(p \omega_1  t),
\label{sppan}
\end{equation}
where we define the decay time
\begin{equation}
t_D\equiv\frac{\omega_1}{ \sigma |e_2|}.
\label{Eq:tD}
\end{equation}
Expression (\ref{sppan}) is valid up to the time when the discrete nature of the spectrum, neglected with the use of the integral, finally manifests itself and induces fluctuations of the survival probability around  its asymptotic value. 

With the  expressions (\ref{sppan}) and %Eq.~
(\ref{IPR}), the equation for the survival probability in (\ref{SPdef}) becomes,
\begin{equation}
SP(t)  \approx \frac{\omega_1}{2\sigma\sqrt{\pi}}
\left\{ 1 +2   \displaystyle\sum_{p=1} 
 \exp \left[-p^2\left( \frac{ \omega_1^2}{4 \sigma^2} +  \frac{t^2}{t_D^2} \right)\right]  \cos(p \omega_1 t) \right\},
\label{SPsuman}
\end{equation}
which is one of the main results of this paper. 
Equation~(\ref{SPsuman}) can also be expressed as a convergent series in terms of the Jacobi theta function ~\cite{Jacobi}, $\Theta_3(x,y)=1+2\sum_{p=1} y^{p^2}\cos(2px)$, using $x=\omega_1 t/2$ and 
$y = \exp \left( - \frac{ \omega_1^2}{4 \sigma^2} -  \frac{t^2}{t_D^2} \right)$,
\begin{equation}
SP(t) \approx \frac{\omega_1}{2\sigma\sqrt{\pi}}\Theta_3(x,y).
\label{theta}
\end{equation}

As one sees from %Eq.~
(\ref{sppan}), the amplitude 
%at $t=0$
 of each component $SP_p(t)$ scales exponentially with $-p^2$. Every $SP_p(t)$ is an oscillating function with frequency $p\,\omega_1 $ modulated in time by a Gaussian function 
\begin{equation}
SP_p^{Decay}(t) =\frac{\omega_1}{\sigma \sqrt{\pi}} \exp \left[ - p^2 \left(\frac{ \omega_1^2}{4 \sigma^2} +  \frac{t^2}{t_D^2}\right) \right]
\label{SPdecay}
\end{equation}
with decay time  $t_{D}^{(p)}=t_D/p$. The decay of the oscillations of the survival probability in %Eq.~
(\ref{SPsuman}) is controlled by the sum of these Gaussians, $SP^{Decay}(t)=IPR+\sum_{p\geq 1}SP^{Decay}_p(t)$.

%%%%%
\begin{figure}
\begin{center}
\begin{tabular}{lc}
\rotatebox{90}{\hspace{100pt} \rotatebox{-90}{(a)}} & \includegraphics[width=.7 \textwidth]{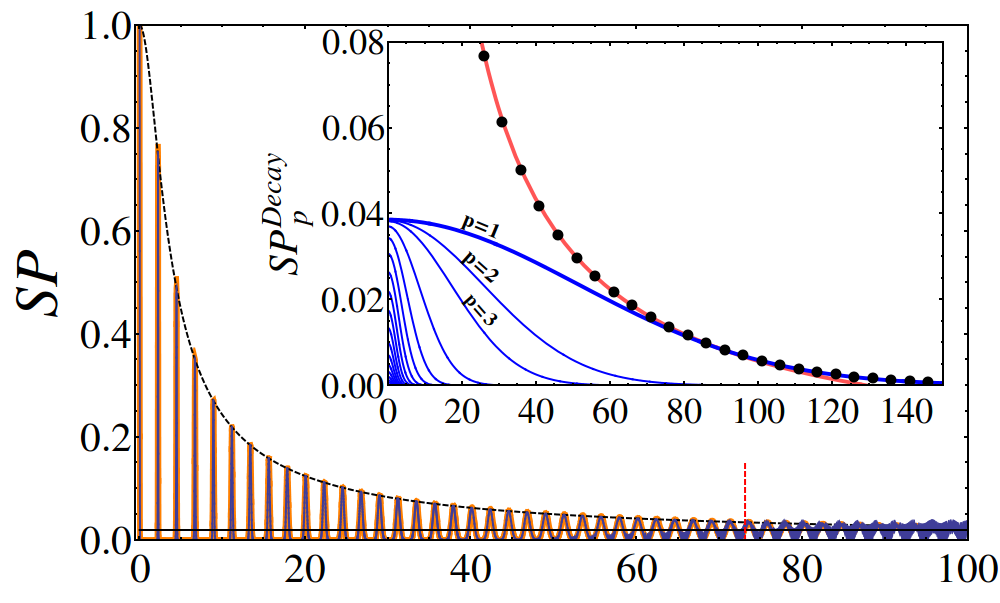} \\%&
\rotatebox{90}{\hspace{70pt} \rotatebox{-90}{(b)}} & \includegraphics[width=.7\textwidth]{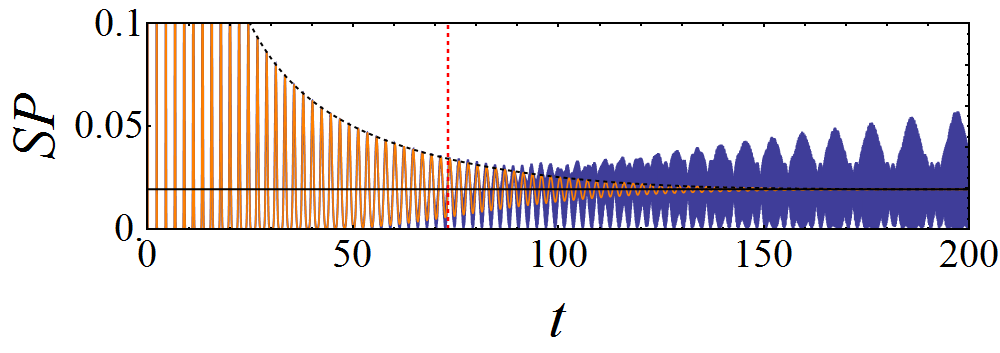}
\end{tabular}
\end{center}
\caption{(a) Survival probability for the same coherent state as in figure~\ref{fig:gaussLMG} obtained numerically (dark blue line) and using the analytical result (light orange line) in %Eq.~
(\ref{SPsuman}). The curves are almost indistinguishable up to $t_D$ (vertical dashed line). The dashed line depicts the analytical decay of the oscillations.   The inset shows the decay of the different $p$-components ($p>0$, bottom curves), their sum (black dots), and the power law fitting $(2.506/t)-IPR$ (solid red line).  (b) is similar to (a), but for longer times. The %vertical dashed line in both main panels indicates the decay time $t_D$ and the 
horizontal black line is the $IPR$.
}
\label{fig:spLMG}
\end{figure}
%%%%%

In the inset of figure~\ref{fig:spLMG} (a), we show the contribution from each $SP_p^{Decay}(t)$. The components with large $p$ decay faster than those with small $p$. At long times, the sum of Gaussians is dominated by the $p=1$ component. Therefore, the decay time of $SP_1(t)$ is also the decay time of the entire $SP(t)$  and is given by $t_D$ from %Eq.~
(\ref{Eq:tD}).  The larger the anharmonicity is, the shorter the decay time becomes, that is faster equilibration. The semi-classical approximation for $t_D$ is obtained from (\ref{limclassom}) and (\ref{e2sc}) as $t_D=2J/(\sigma|d\omega_{cl}/d\epsilon|)$.

We emphasize that in this one-degree of freedom case, only three parameters are needed to fully describe the survival probability at any time up to the equilibration time. As seen from %Eq.~
(\ref{SPsuman}), they are the energy  width $\sigma$ which can be calculated analytically (see \ref{App2})%of the Gaussian function that describes the dependence in energy of the components $|c_k|^2$ of the initial state
, the mean energy separation between eigenenergies $\omega_1$ (approximated by $\omega_{cl}$  in the semi-classical limit), and the anharmonicity in the energy spectrum $e_2$ [with semi-classical limit in (\ref{e2sc})]. These parameters depend 
on the initial state,  and we have tested  the ability of our  analytical expression (\ref{SPsuman}) to describe the numerical $SP$  for many different  states, founding a remarkable agreement (as in the case shown below), provided  the initial state is far  from critical energies.

In  \ref{AppPar}, we show   the dependence of $\sigma$, $\omega_1$,  $e_2$ and  the decay time (\ref{Eq:tD})  on the coordinates of the initial coherent state. Likewise, the small regions in  the coherent parameter space close to the critical energies  where our approach fails, are identified for the considered case with $J=2000$.

%%%%%%%%%%%%%%%%%%%%%%%%%%%%%%%%%%%%%%%%
\subsection{Comparison with Numerics}
\label{Subsec:CompNum}

In the main panels of figure~\ref{fig:spLMG}, we compare the analytical expression (\ref{SPsuman}) and the numerical results for the LMG model using the same parameters and initial state as in figure~\ref{fig:gaussLMG}. The relevant parameters obtained with (\ref{components}) [or (\ref{eq-app2:sigLMG})], (\ref{w1es}), and (\ref{e2es}) are $(\sigma,\omega_1, e_2)=(41.08,2.82,-9.38\times 10^{-4})$, which gives the decay time  $t_D=73.09$.% (indicated with vertical dashed lines in the main panels of figure~\ref{fig:spLMG}).

The analytical approximation reproduces remarkably well the numerical results up to $t_D$. The two lines in the main panel of figure~\ref{fig:spLMG} (a) can hardly be distinguished. The numerical oscillations as well as their decay  agree extremely well with the analytical expression (\ref{SPsuman}).% are in excellent agreement with the numerics, as well as their decay $SP^{Decay}(t)$, indicated with the dashed line  and given by  %Eq.~
%(\ref{SPsuman}) without the cosine function.

At times of the order of $t_D$, the decay of the oscillations of the survival probability is power law, in accord with %mentioned in %Ref.~
\cite{Santos2016}. % The power-law decay, fitted as
This is confirmed with the fit  $2.506/t$  illustrated with a solid red line in the inset of figure~\ref{fig:spLMG}  (a). This behavior, including the pre-factor, can be justified analytically in the semi-classical limit (see the next subsection and \ref{App6bis} %Section H in \cite{SupMat}
). 

Figure~\ref{fig:spLMG} (b) makes more evident what happens at long times,  when the discrete nature of the spectrum becomes important. Beyond $t_D$, the numerical curve fluctuates around the infinite time average, while the analytical expression simply stabilizes at $IPR$.

%%%%%%%%%%%%%%%%%%%%%%%%%%%%%%%%%%%%%%%%
\subsection{Classical Limit}
\label{sec:ClassicalLimit}

The  analytical expression (\ref{SPsuman}) for the survival probability %and its parameters have
has a well defined classical limit.
In this   limit, $e_2\approx f_e/J$ [see (\ref{e2sc})] goes to zero faster  than the growth of $\sigma\approx f_\sigma\sqrt{J}$ [see (\ref{sigma})]%, shown in \cite{Schliemann2015} and in (\ref{sigma})
. Consequently, the decay time goes to infinity,
\begin{equation}
\lim_{J\rightarrow\infty } t_D= \lim_{J\rightarrow\infty }\frac{\omega_1}{ \sigma |e_2|}\propto\lim_{J\to\infty} \sqrt{J}=\infty,
\end{equation}
 and the expression (\ref{SPsuman}) for the $SP$ becomes a sum of Kronecker deltas (see \ref{App6bis})
 \begin{equation}
\lim_{J\rightarrow\infty}SP(t)= \sum_{n\in\mathbb{Z}}\delta_{t,n\tau}f_n, 
 \end{equation}
  with $\tau=2\pi/\omega_{cl}$ and $f_n= (1+ (4 \pi f_\sigma^2 f_e/\omega_{cl}^3)^2 n^2)^{-1/2}$%(1+ f_\sigma^4 (d^2I/d\epsilon^2)^2 n^2 )^{-1/2}$
. This result indicates periodic instantaneous revivals, which are indeed  % is the expected for a regular classical system, not so the  decay of their amplitude, even if it is power-law.   the reason for the   power-law decay of their amplitudes is  .%     This
% are %indeed 
% an 
 expected for the survival probability in a one-degree of freedom, regular classical system.

%%%%%
\begin{figure}
\begin{tabular}{lccc}
 &(a) & (b) & (c) \\
% \scalebox{1.1}{$|c_k|^2$}\ &\hskip -20 pt\includegraphics[width=.29\textwidth]{fig3a}
%%&\hskip -20 pt\includegraphics[width=.29\textwidth]{fig3b}
%%&\hskip -20 pt\includegraphics[width=.29\textwidth]{fig3c}
%%\\&(d) & (e) & (f) \\
%%      &\hskip -12 pt\includegraphics[width=.31\textwidth]{fig3d}
%%      &\hskip -12 pt\includegraphics[width=.31\textwidth]{fig3e}
%%      &\hskip -12 pt\includegraphics[width=.31\textwidth]{fig3f}
\scalebox{1.1}{$|c_k|^2$}\ &\hskip -12 pt\includegraphics[width=.31\textwidth]{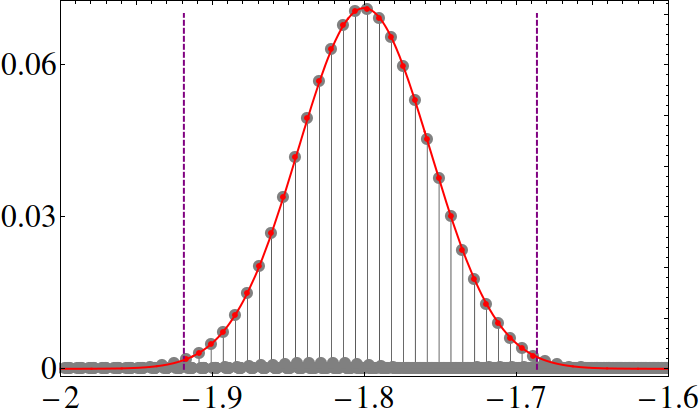}
&\hskip -12 pt\includegraphics[width=.31\textwidth]{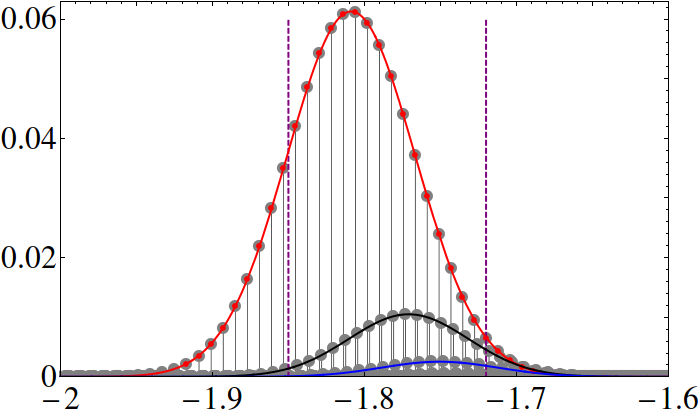}
&\hskip -12 pt\includegraphics[width=.31\textwidth]{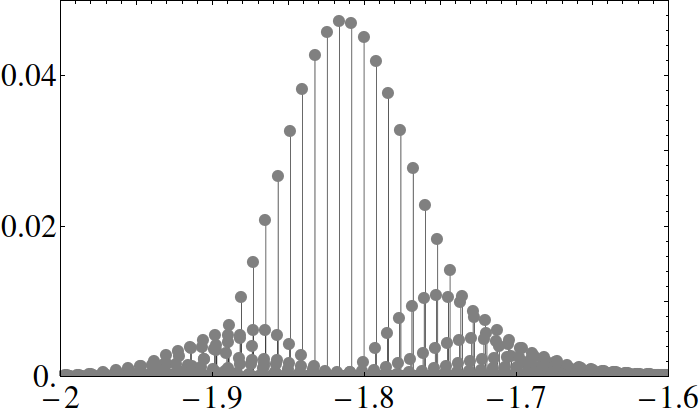}
\\&(d) & (e) & (f) \\
      &\hskip -12 pt\includegraphics[width=.31\textwidth]{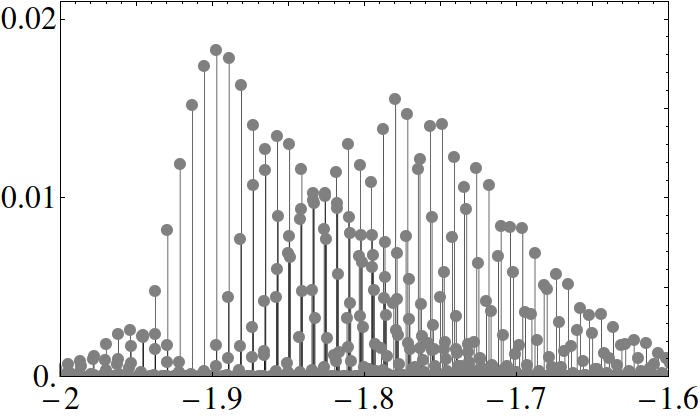}
      &\hskip -12 pt\includegraphics[width=.31\textwidth]{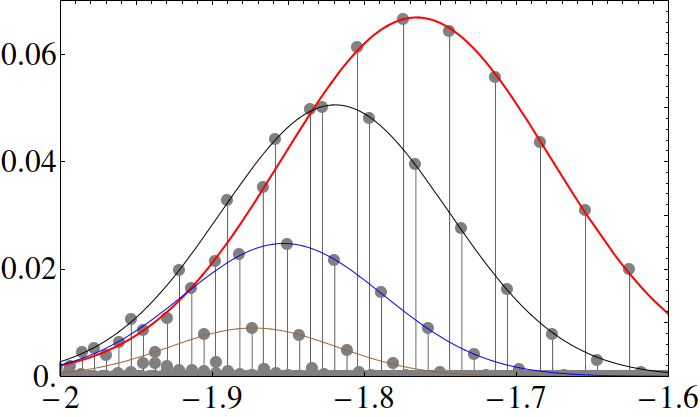}
      &\hskip -12 pt\includegraphics[width=.31\textwidth]{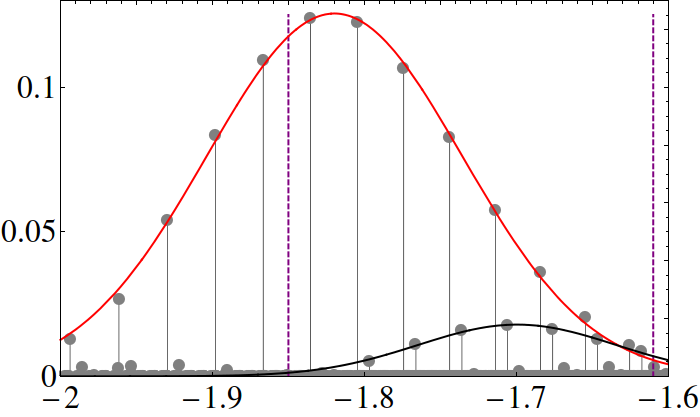}
      \\
& $E_{k}/J$ & $E_{k}/J$ & $E_{k}/J$
\end{tabular}
\caption{Eigenenergy components of a sample of coherent states with the same mean energy ($E/J=-1.8$) for the Dicke model with $\omega=\omega_0=\gamma=1$ and $J=120$.  The Bloch coherent parameters are $\phi_o=0$ and 
 $j_z=-0.505$ (a), $-0.452$ (b),  $-0.116$ (c), $0.019$ (d), $0.113$ (e), $0.140$ (f), and the Glauber parameters are $p_o=0$ and $q_o$ given by the condition $\langle  \alpha_o z_o|H_{D}|\alpha_o z_o\rangle/J=-1.8$. Solid lines in panels (a), (b) and (f) represent Gaussian fits (parameters in table~\ref{parameters}) for the sub-sequences with the largest components $|c_k|^2$. Vertical dashed lines in panels (a), (b) and (f) indicate the sample of energy components used in figures \ref{fig:SPdicke}(a) and \ref{fig:disga2}.}
\label{fig:dis}
\end{figure}
%%%%%

With the  asymptotic expressions for $\sigma$,  $\omega_1$ and  $e_2$, it is possible to justify the power law observed in figure~\ref{fig:spLMG} for the decay of the survival probability  at times of the order $t_D$. For this,  we investigate  $SP^{Decay}(t)$, that is, %Eq.~
(\ref{SPsuman}) without the cosine function. 
%Section H in \cite{SupMat}
\ref{App6bis} shows  that for large $J$, $SP^{Decay}\approx  c/t$, where $c$ is an asymptotically  finite value given by $c=\frac{\omega_{cl}^2}{2 \sigma^2 |e_2|}$. 
 For the parameters used in figure~\ref{fig:spLMG}, we find $c=2.512$,  which is in excellent agreement with the fit in the inset of figure~\ref{fig:spLMG} (a), which gives $c=2.506$.
 
%%%%%%%%%%%%%%%%%%%%%%%%%%%%%%%%%%%%%%%%
%%%%%%%%%%%%%%%% DICKE MODEL% %%%%%%%%%%%%%%%

\section{Survival probability in two-degree-of-freedom models}
\label{Sec:DMSP}

We now use the Dicke model in the superradiant phase to characterize the dynamics of quantum models with two-degrees of freedom. 
 The Dicke model has both regular and chaotic regimes. The classical regular dynamics occurs at low energies~\cite{Bastarrachea14PRA-2,Chavez2016} and is accounted for by quasi-integrals of motion~\cite{Relano16,BastaJPA17}. %%In the quantum domain, for the different subspaces associated with the quantum numbers of the quasi-integrals of motion, the  spectrum probed by   coherent states at large $J$ is quasi-harmonic. This is not the case in the chaotic regime.

The description of the evolution of the survival probability for the Dicke model is richer than what we 
%find for 
can find in models with one-degree of freedom. This happens because, in general, the projection of coherent states into the energy eigenbasis no longer leads to a single sequence of components $|c_k|^2$ following a single Gaussian function, as  for the LMG model in figure~\ref{fig:gaussLMG}~(a). Instead, the components of the initial state now form different sub-sequences (figure ~\ref{fig:dis}). In the regular regime, these sub-sequences are overall still represented by Gaussian functions, but of different means and widths. These various sub-sequences interfere and lead to a more complex behavior of the survival probability. The energy eigenbasis decomposition of the coherent states is closely related with the properties of the classical phase space of the Dicke model, as it will be shown below. %To better understand this relationship, we compare the structures of different coherent states written in the energy eigenbasis [figure~\ref{fig:dis}] with their location in the classical phase space [figure~\ref{fig:cs} (b)]. 

After a discussion in subsection \ref{Sec:StatesDicke} about the different energy distributions of the coherent states displayed in figure \ref{fig:dis}, we select four representative cases and analyze their survival probability in the following subsections. We start in Sec.~\ref{Sec:oneDicke} with the state in figure \ref{fig:dis}~(a). This initial state activates only one of the  degree of freedom, and, as a  consequence,  the analytical expression of the previous section describes very well the numerical results. %,  analogously to what we found for the LMG model.
 In section~\ref{Sec:Interference}, we consider the states from figures \ref{fig:dis}~(b) and (f).  They are representative cases of initial states activating  simultaneously  the two-degrees of freedom, which  yields to interference terms that  we are able to describe analytically with formula (\ref{Eq:incon}). For Sec.~\ref{Sec:Resonance}, we choose the state from figure \ref{fig:dis}~(d) to illustrate the effects  of the non-linear instabilities and unveil the signature of classical chaos in  the quantum dynamics.  The  selected initial states are representative of the whole cases that can be found in the regular regime of the Dicke model.  In \cite{LermaAIP18}   other coherent  states at the same and also  larger energies than the one used here,  are studied, reinforcing  the validity of the results presented and  discussed below.

\subsection{Initial coherent states}
\label{Sec:StatesDicke}

\begin{figure}
\begin{tabular}{lcc}
 & (a) & (b)\\
\rotatebox{90}{\hspace{80pt} \rotatebox{-90}{\scalebox{1.2}{$j_z$}}}& \includegraphics[width=.44\textwidth]{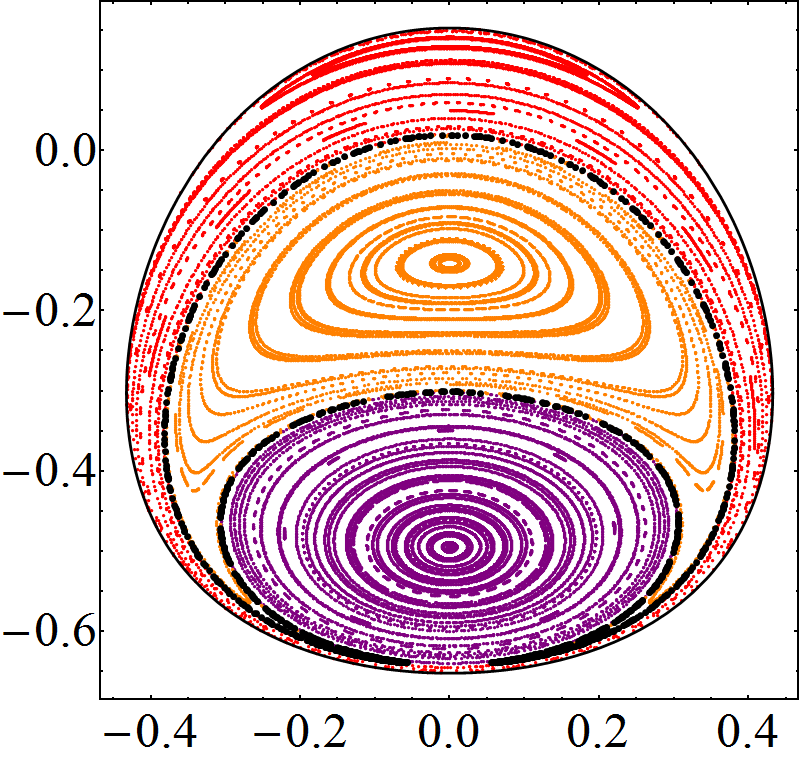} & \includegraphics[width=.44\textwidth]{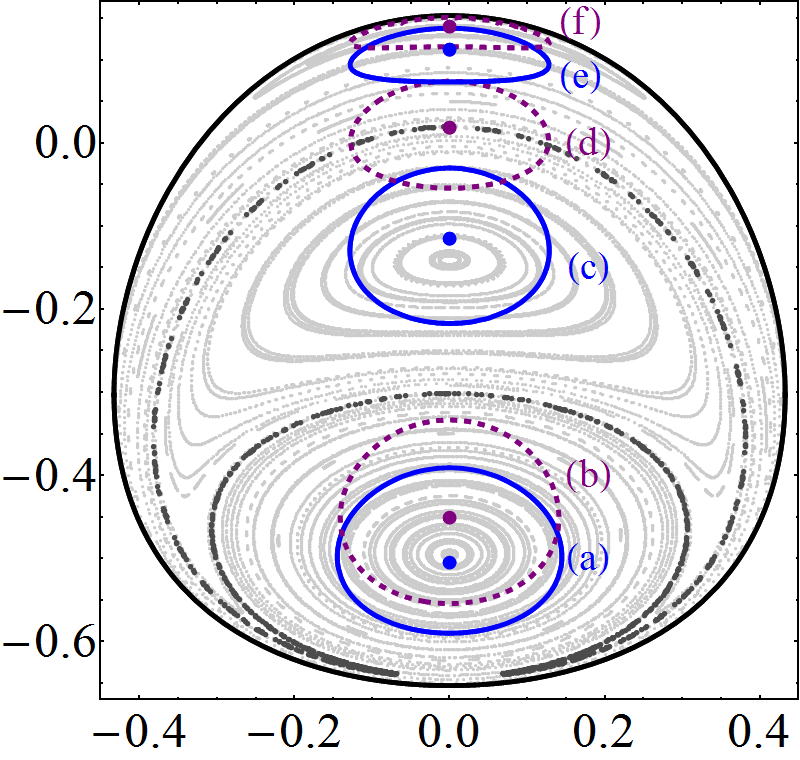}\\
  &\ \  \scalebox{1.2}{$\phi$}&\ \  \scalebox{1.2}{$\phi$}
\end{tabular}
\caption{ (a) Poincar\'e sections ($p=0$) projected in the plane $j_z-\phi$ for the classical Dicke model with the same parameters and energy as figure~\ref{fig:dis}. The dark black trajectory indicates the separatrix between the region of nonlinear resonances (central light orange trajectories) and the regions of the  adiabatic modes (outer light red trajectories for the pseudospin  and dark purple for the bosonic mode).  (b) Location of the coherent states from figure~\ref{fig:dis} in the classical phase space. Dots (from bottom to top) indicate the coherent states from figure~\ref{fig:dis} (a) to figure~\ref{fig:dis} (f). The closed curves that encircle the dots represent the spreading of the corresponding wave functions (level curves $|\langle z,\alpha|z_o,\alpha_o\rangle|^2=e^{-1}$) for $J=120$.  }
\label{fig:cs}
\end{figure}

In figure~\ref{fig:dis}, we fix $\phi=0$ and $p=0$, vary $j_z$, and determine $q$ from the condition that guarantees that  all chosen coherent states have the same mean energy $E/J=-1.8$, which is relatively close to the ground-state ($E_{GS}/J=-2.125$).
Regular dynamics dominates this energy region, as seen in figure~\ref{fig:cs} (a). This figure shows Poincar\'e sections for the classical limit of the Dicke model at $E/J=-1.8$. The closed loops, covering  the whole Poincar\'e surface, reflect the existence of invariant tori. Their nature can be revealed in light of the adiabatic approximation \cite{Relano16,BastaJPA17}: for the parameters  ($\omega=\omega_0=\gamma=1$) and energy chosen here, the dynamics of the bosonic variables $(p,q)$  is slower than that of the  pseudospin variables. The pseudospin precesses rapidly around a slowly changing $q$-dependent axis. The nearly constant angle $\beta$ that forms the  pseudospin with respect to the precession axis defines an effective one-dimensional adiabatic potential for the bosonic variables. If the angle $\beta$ is small, the amplitude of the bosonic variables is large and vice-versa.%, if the angle $\beta$ is large the amplitude of the bosonic variables is small.

The Poincar\'e sections in figure~\ref{fig:cs} (a) can then be understood as follows:
\begin{itemize}
\item The trajectories rotating  around $(j_z,\phi)\approx (-0.5,0)$ (plotted in purple) correspond to small precessing angles $\beta$ and wide amplitudes of the bosonic excitations.
\item The trajectories rotating around $(j_z,\phi)\approx (0.15,0)$ (plotted in red) have large $\beta$ and consequently small displacements of the bosonic variables.
\item The trajectories in the center (plotted in orange), rotating around the point  $(j_z,\phi)\approx (-0.15,0)$,   indicate the breaking of the adiabatic approximation. They emerge from nonlinear resonances between the adiabatic modes. These trajectories are the precursors of ample chaotic regions that appear for energies larger than the one considered here. In fact, a detailed view of the separatrix between this last set of trajectories and the two former ones reveals the existence of a narrow region with classical chaotic trajectories~\cite{BastarracheaPS2017}.
\end{itemize}

The six coherent states of figure~\ref{fig:dis} sample the three classical regions listed above.
In figure~\ref{fig:cs}~(b), we show where these  states  fall in the Poincar\'e surface. Each point in  figure~\ref{fig:cs} (b) indicates the phase-space coordinates associated with the coherent state parameters ($z_o, \alpha_o$), and the curve surrounding each point represents the spreading of the corresponding coherent state wave function in phase space (level curves $|\langle z,\alpha|z_o,\alpha_o\rangle|^2=e^{-1}$ for $J=120$).

%%%%%
%%%%%

According to the list above, the states in figures.~\ref{fig:dis} (a) and (b) are associated with large bosonic amplitudes and those in  figures.~\ref{fig:dis} (e) and (f)  with large pseudospin precession angles. They have one sequence (a) or sub-sequences (b, e, f) of components described by Gaussian distributions.
The differences of consecutive energies in the sub-sequences of figures.~\ref{fig:dis}  (e) and (f) are larger than in figures.~\ref{fig:dis}  (a) and (b). This can be qualitatively understood from the classical model, because the pseudospin has  faster dynamics than the bosonic variables, and thus larger oscillation frequencies. 

The state in figure~\ref{fig:dis} (a) is representative of nearly pure bosonic excitations. Its components $|c_k|^2$ are well approximated by a single Gaussian function.  This situation is equivalent to what we have for the LMG model, so the analytical expression (\ref{SPsuman}) is still applicable here.  This state corresponds classically to  pseudospin precessing angle $\beta=0$ and maximal amplitude of the bosonic variables.

In contrast, the state in figure~\ref{fig:dis} (f) is representative of nearly pure pseudospin excitations.  Its components $|c_k|^2$ are well described by a dominant  Gaussian distribution with  a second smaller  Gaussian sub-sequence. In the classical picture, this state corresponds  to nearly maximal  precessing pseudospin angle  and nearly zero amplitude of the bosonic variables.

The states in figures.~\ref{fig:dis} (b) and (e) have more than a single sequence of components; three Gaussians are identifiable in (b), while four are distinguished in (e). The presence of several sub-sequences of components in these states corresponds classically to the simultaneous excitation of different adiabatic modes, with the dominance of one of them, the bosonic one in figure~\ref{fig:dis}~(b) and the pseudospin mode in figure~\ref{fig:dis}~(e).

The state in figure~\ref{fig:dis} (c), located close to the center of the region of nonlinear resonances, exhibits a dominant Gaussian sub-sequence and many smaller ones, while the coherent state in figure~\ref{fig:dis} (d)  has a complicated structure with so many eigenstates participating that it is hard to identify the sub-sequences (if any).  In the classical phase space of  figure~\ref{fig:cs} (b), this state is located in the  unstable separatrix between the  region of non-linear resonances and the fast mode of the adiabatic approximation, 
% region of trajectories with wide pseudospin precessing angle, 
where, as it is known \cite{ReichlBook},   classical chaos emerges.

%%%%%%%%%%%%%%%%%%%%%%%%%%%%%%%%%%%%%%%%
\subsection{One-sequence coherent state}
\label{Sec:oneDicke}
%%%%%
\begin{figure}
\begin{tabular}{cc}
 (a)&(b)\\
\includegraphics[width=.45\textwidth]{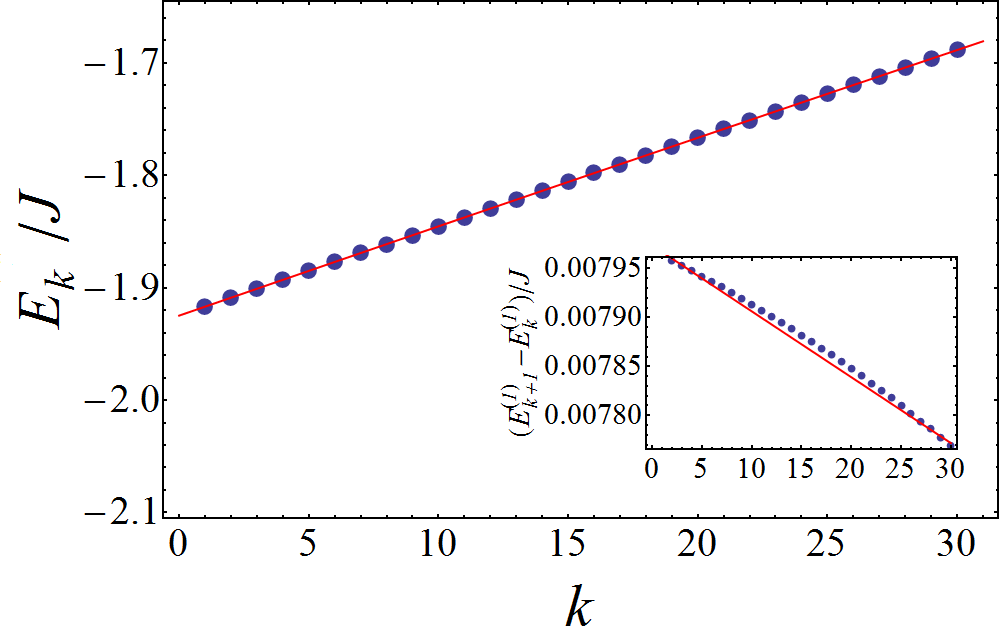}&\includegraphics[width=.45\textwidth]{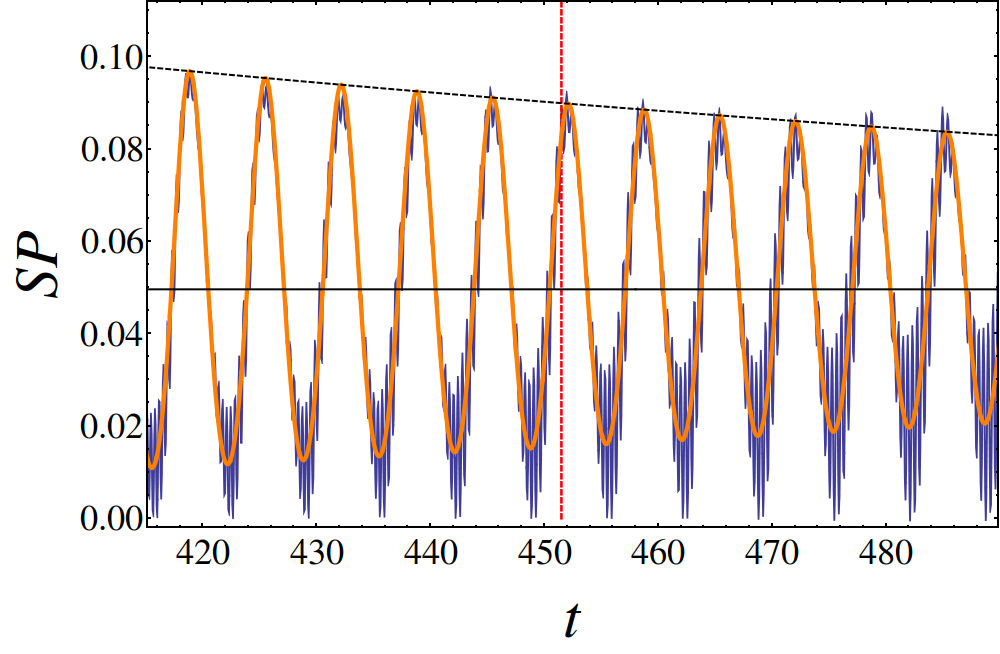}\\%\rotatebox{90}{\hspace{20pt}\rotatebox{-90}{\includegraphics[width=.45\textwidth]{fig5c}}}\\
 \multicolumn{2}{c}{(c)}  \\ 
 \multicolumn{2}{c}{\includegraphics[width=.7\textwidth]{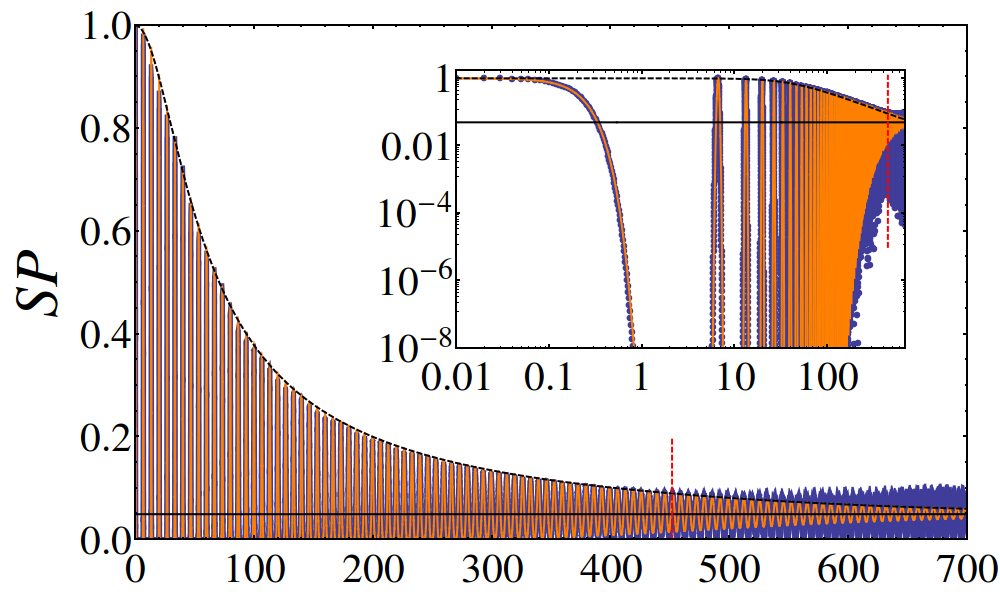}} \\ 
 \multicolumn{2}{c}{\ \ \ \ \ \  \ \scalebox{1.3}{$t$} }  
 \end{tabular}
\caption{ (a) Energies (dots) of the 30 largest components of the coherent state  of figure~\ref{fig:dis}~(a) (the ones located  between the two vertical dashed lines in that figure). The solid line is a fit using %Eq.~
(\ref{principalspectrum}). The inset shows the difference between consecutive energies and the corresponding fit. (c) Survival probability for the coherent state of figure~\ref{fig:dis}~(a): numerical curve (dark blue) and analytical expression (light orange). The inset shows the same figure in  log-log scale. (b) Closer view around the decay time $t_D$. In both panels, the dashed black line depicts the analytical decay of the oscillations of $SP$.  The vertical dashed line indicates the decay time $t_D=451.5$ and the horizontal solid black line is the $IPR=0.0496$.}
\label{fig:SPdicke}
\end{figure}

In figure~\ref{fig:dis}~(a), we show a Gaussian fit to the energy components of the coherent state. The mean and the width $\sigma$ obtained from the fitting match those calculated analytically through the expectation values $\langle H_{D} \rangle$ and $\langle H_D^2\rangle$ (see \ref{App2}). This agreement confirms that this state is indeed very well described by a single sequence of energy components.

The energy levels $\{ E_k\}$ that are relevant to the evolution of the coherent state, {\em i.e.} those with non-negligible $|c_k|^2$, are  very well  described  by  the semi-classical approximation %Eq.~
(\ref{principalspectrum}), as can be seen in figure~\ref{fig:SPdicke}~(a). A tiny discrepancy is visible by plotting the energy difference $E_{k+1} - E_k$ in the inset of figure~\ref{fig:SPdicke}~(a), which could be related with
the small $J$ accessible to our numerical analysis of the Dicke model ($J=120$). However, as we show below, %Eq.~
(\ref{principalspectrum}) can still be successfully employed for the description of the survival probability. 

The  analytical expression (\ref{SPsuman}) used for the LMG model can be used here also. 
Using (\ref{w1es}) and (\ref{e2es}) in the numerically evaluated spectrum, we obtain 
$(\omega_1,e_2)=(0.9456,-0.399\times 10^{-3})$, which together with the calculated width $\sigma/J=0.0436$ give the decay time $t_D=451.5$. 

The analytical approximation (\ref{SPsuman}) and the numerical results for the survival probability are compared in figures.~\ref{fig:SPdicke} (b) and (c) both in linear (main panels) and log-log (inset) scales. The analytical approximation gives a very accurate description of $SP(t)$ from $t=0$ until $t_D$. 
Beyond the equilibration time, the  discreteness of the energy spectrum becomes relevant. It leads to small fluctuations that are not captured by the analytical expression, as seen in figure~\ref{fig:SPdicke}~(b).

%%%%%%%%%%%%%%%%%%%%%%%%%%%%%%%%%%%%%%%%
\subsection{Interference terms}
\label{Sec:Interference}

When the components of the initial state can be fitted with more than a single Gaussian, as in figure~\ref{fig:dis} (b) and figure~\ref{fig:dis} (f), we use the index $i$ to denote the components $|c_k^{(i)}|^2$, energies $\{E_k^{(i)}\}$, and the Gaussian curve $g_k^{(i)}=A_i e^{-(E_k^{(i)}-\bar{E}_i)/(2 \sigma_i^2)}$ associated with each sub-sequence. Three Gaussians ($i=1,2,3$) are used for the state in figure~\ref{fig:dis} (b) and two ($i=1,2$) for the state in figure~\ref{fig:dis} (f). In these cases, %Eq.~
(\ref{SPdef}) for the survival probability can be written as
\begin{align}
SP(t)=\left|\sum_{ik} |c_k^{(i)}|^2 e^{-i E_k^{(i)}t}\right|^2 
%&= \sum_{i} SP^{(i)}(t)+\sum_{i<j}\sum_{kk'} 2 |c_k^{(i)}|^2|c_{k'}^{(j)}|^2\cos\left[\left(E_{k'}^{(j)}-E_{k}^{(i)}\right)t\right]
%\nonumber \\&
=\sum_{i} SP^{(i)}(t)+\sum_{i<j} SP_I^{(ij)}(t).
\label{Eq:SPwithInt}
\end{align}
The novelty with respect to (\ref{SPsuman}) is the interference terms $SP_I^{(ij)}$.
The  steps involved in the derivation of the terms $SP^{(i)}$ are similar to those taken in Sec.~\ref{Sec:LMG} and  an equation equivalent to %Eq.~
%(\ref{SPsuman})
(\ref{theta}) is obtained 
%%\begin{align}
%%&SP^{(i)}(t) = \frac{A_i^2 \sigma_i\sqrt{\pi}}{\omega_1^{(i)}}   \left\{\!\! 1+2 \! \displaystyle \sum_{p=1}  \!  \exp\left[-p^2\left( \frac{\left(\omega_1^{(i)}\right)^2}{4 \sigma_i^2} + \frac{t^2}{\left(t_D^{(i)}\right)
%%^2}  \right)\right]  \cos(p\omega_1^{(i)} t) \!\!  \right\}. 
%%\label{Eq:iscon} 
%%\end{align}
\begin{equation}
SP^{(i)}(t) = \frac{A_i^2 \sigma_i\sqrt{\pi}}{\omega_1^{(i)}}\Theta_3(x_i,y_i), 
\label{Eq:iscon}
\end{equation}
with  $x_i= \omega_1^{(i)} t/2$  and $y_i= e^{-\left( \omega_1^{(i)}/2 \sigma_i\right)^2} e^{ - \left(t/t_D^{(i)}
  \right)^2}$.
%By fitting the sub-sequences of components with Gaussian functions, we can obtain the mean energy $\bar{E}_i$, the width $\sigma_i$, and the amplitude $A_i$. 
Analogously to Sec.~\ref{Sec:LMG}, the decay time of each isolated sub-sequence is
$ t_D^{(i)}= \omega_1^{(i)}/(|e_2^{(i)}|\sigma_i) $; the frequency $\omega_1^{(i)}=E_{kmax+1}^{(i)}-E_{kmax}^{(i)}$ is the  difference of the closest energies of the $i$-th sub-sequence to the mean energy $\bar{E}_i$, with $E_{kmax}^{(i)}\leq \bar{E}_i \leq E_{kmax+1}^{(i)}$; and  the anharmonicity is $e_2^{(i)}= (E_{kmax+1}^{(i)}+E_{kmax-1}^{(i)})/2-E_{kmax}^{(i)}$.

%%%%%
\begin{figure}
\begin{tabular}{cc}
(a)& (b)\\
  \includegraphics[width=.47\textwidth]{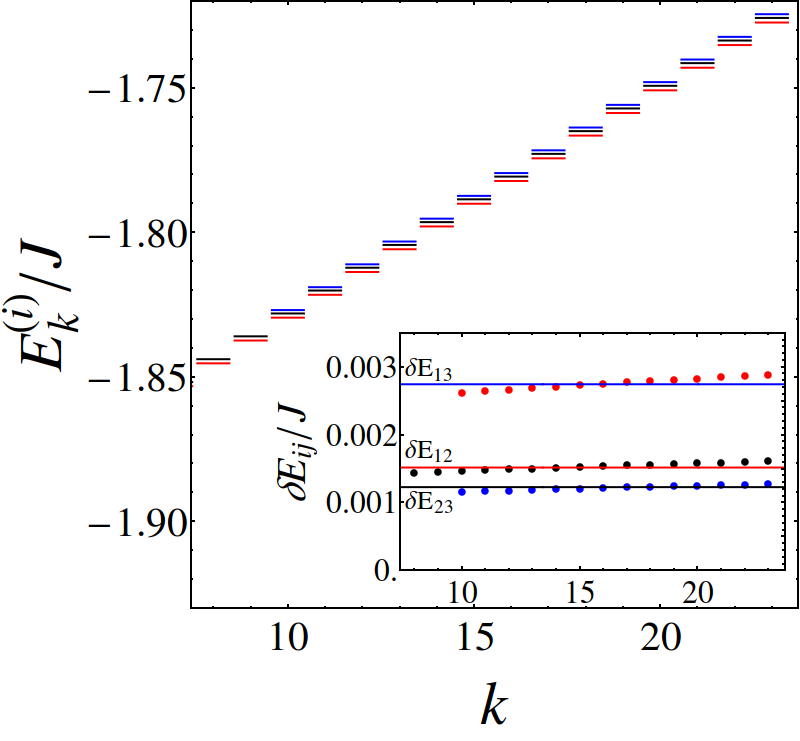}
  &  \includegraphics[width=.47\textwidth]{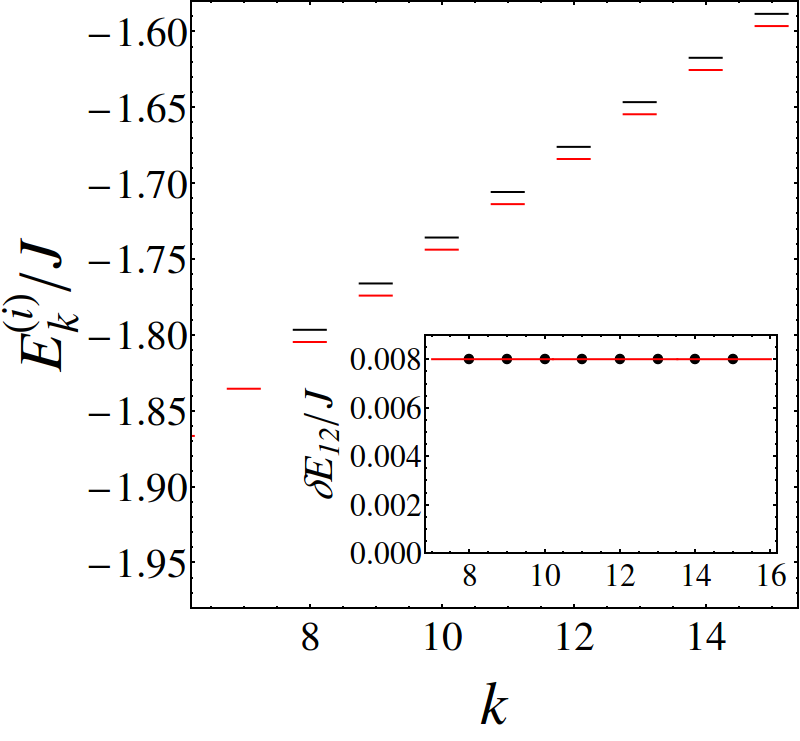}
 \end{tabular}
\caption{ (a) Energies of the largest components of the three sub-sequences contributing to the coherent state in figure~\ref{fig:dis}~(b).  The sampled energies are the ones located between the two vertical dashed lines in that figure. Each sub-sequence has a different color, which corresponds  with the colors used in figure~\ref{fig:dis}~(b) . The inset shows the energy differences between two distinct sub-sequences (dots) and the respective mean values (horizontal lines).  (b) Similar to panel (a), but for the  largest components of the two principal sub-sequences of the coherent state in figure~\ref{fig:dis}~(f). }
\label{fig:disga2}
\end{figure}
%%%%%

To obtain an analytical expression for $SP_I^{(ij)}$, we use the same strategy used in Sec.~\ref{Sec:LMG}, namely we separate the terms according to the index distance $p$ between the eigenvalues %(see \ref{App8} %Section I of \cite{SupMat}
% for a detailed derivation) 
\begin{equation}
SP_I^{(ij)}(t) = 2 \displaystyle\sum_{p\in\mathbb{Z}} \sum_{k}  |c_k^{(i)}|^2 |c_{k+p}^{(j)}|^2 \cos\left[ (E_{k+p}^{(j)}-E_k^{(i)})t\right].
\label{spiap}
\end{equation}
In addition, we assume that the energy sub-sequences are of the form (\ref{principalspectrum})  and related by a constant shift  $\delta E_{ij}$
\begin{equation}
\delta E_{ij} =E_{k}^{(i)} - E_{k}^{(j)}.
\label{shift}
\end{equation} 
This is an important step in the derivation of an expression for $SP_I^{(ij)}(t)$. 
In figure~\ref{fig:disga2} 
(a) and figure~\ref{fig:disga2} (b), we show the energies $\{E_k^{(i)}\}$ of each Gaussian sub-sequence of the coherent states from figure~\ref{fig:dis} (b) and figure~\ref{fig:dis} (f), respectively. %The energies are obtained numerically, although they could also be estimated from a semi-classical analysis.
 The insets of figure~\ref{fig:disga2} (a) and (b) confirm the validity of %Eq.~
 (\ref{shift}). The agreement is not perfect, but it should improve for larger $J$.

%%%%%
%%%%%
\begin{table}[b]
\begin{tabular}{|c|c|c|c|c|c|c|| c|c|c|}
\hline
$i$ & $A_i$& $\bar{E}_i/J$ & $\sigma_i/J$& $\omega_{1}^{(i)}/J$ & $e_{2}^{(i)}/J$& $t_D^{(i)}$ & $i,j$  &$\omega_{ij}/J$  & $\delta E_{ij}/J$\\ 
\hline
\multicolumn{10}{|c|}{Coherent state of figure~\ref{fig:dis}(b)}\\
\hline 
1 & 0.0612 & -1.809  & 0.0421 & 0.00788 &$ -3.3\times 10^{-6}$ & 476.80 & 1,2 & 0.00786 & 0.00152\\
2 & 0.0105 & -1.771  & 0.0394 & 0.00786 &$ -3.7\times 10^{-6}$ & 447.08  & 1,3 & 0.00785 & 0.00275\\  
3 & 0.0025 & -1.751  & 0.0382 & 0.00785 &$ -4.1\times 10^{-6}$ & 420.97  &  2,3 & 0.00785 & 0.00123\\
\hline
\multicolumn{10}{|c|}{Coherent state of figure~\ref{fig:dis}(f)}\\
\hline 
1 & 0.126 & -1.820 &  0.0841 & 0.0308 & -0.000135 & 22.57 & 1,2  & 0.0303 &  0.00802 \\
2 & 0.018 & -1.699 &  0.0667 & 0.0297 & -0.000134 & 27.73 & &  & \\
\hline
\end{tabular}
\caption{Parameters determined from the numerical spectrum of the coherent states of figure~\ref{fig:dis} (b) and figure~\ref{fig:dis} (f), and  used in the analytical expression (\ref{Eq:SPwithInt}) for the survival probability plotted in figures \ref{fig:SPin1} and \ref{fig:SPin2}~(a).
}
\label{parameters}
\end{table}
%%%%% 

From the  assumption (\ref{shift}) and using $ |c_k^{(i)}|^2\approx g_{k}^{(i)}$ %=A_i\exp\left[-\frac{(E_{k}^{(i)}-\bar{E}_i)^2}{2 \sigma_i^2}\right]
, the following expression is obtained   for the interference terms (see \ref{App8} %Section I of \cite{SupMat} 
for a detailed derivation),
\begin{align}
&SP_I^{(ij)}(t)=  \frac{2 A_i A_j\sqrt{2\pi}\sigma_i\sigma_j}{\omega_{ij}\sqrt{\sigma_i^2+\sigma_j^2}} 
\displaystyle\sum_{p\in \mathbb{Z}} e^{-\frac{(p \omega_{ij}+\delta E_{ij}+\bar{E}_i-\bar{E}_j)^2}{2(\sigma_i^2+\sigma_j^2)}}e^{-\frac{(\sigma_{ij} p t)^2}{2}}\cos[(\delta E_{ij}+p\omega_{ij})t].
\label{Eq:incon}
\end{align}
Above,
$$
\sigma_{ij}=\frac{2 |e_2^{(i)}| \sigma_i\sigma_j}{\omega_{ij} \sqrt{\sigma_i^2+\sigma_j^2}},
$$
and $\omega_{ij}=E_{k_I+1}^{(i)}-E_{k_I}^{(i)}$ is the energy difference between the eigenvalues of the $i$-th sub-sequence that are closest to the value $E^{(I)}_{ij}$ that maximizes the product of Gaussians $g_{k}^{(i)} g_{k}^{(j)}$. This value is given by   
$$
E^{(I)}_{ij}=\frac{\bar{E}_i \sigma_j^2 + \bar{E}_j \sigma_i^2}{\sigma_i^2 + \sigma_j^2}
$$
and satisfies $E_{k_I}^{(i)}\leq E^{(I)}_{ij}\leq E_{k_I+1}^{(i)}$.

\begin{figure}
\begin{center}
\begin{tabular}{lc}
\rotatebox{90}{\hspace{110pt}\rotatebox{-90}{(a)}} &\includegraphics[width=.7\textwidth]{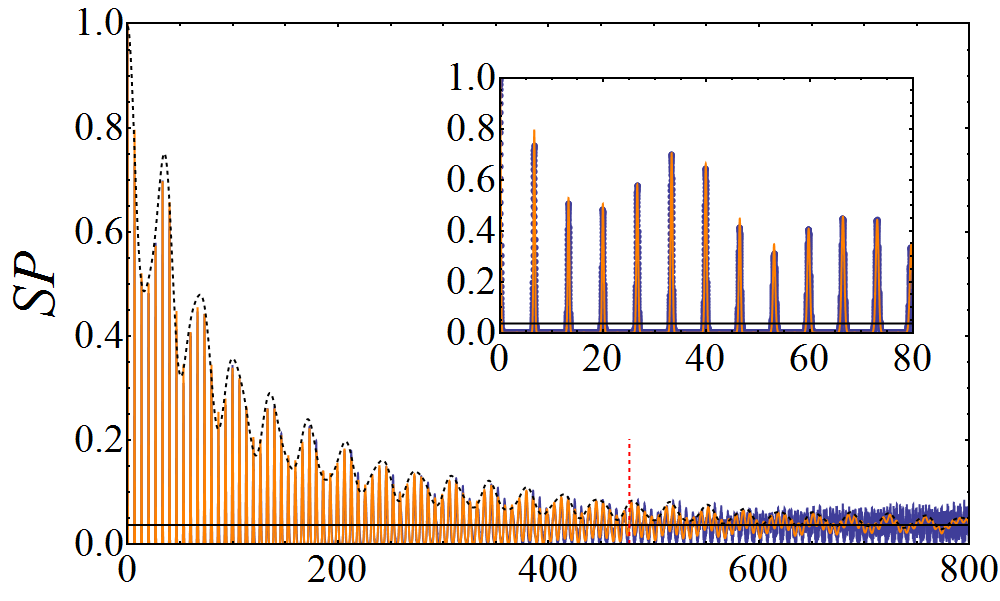}\\
\rotatebox{90}{\hspace{80pt}\rotatebox{-90}{(b)}} &  \includegraphics[width=.7\textwidth]{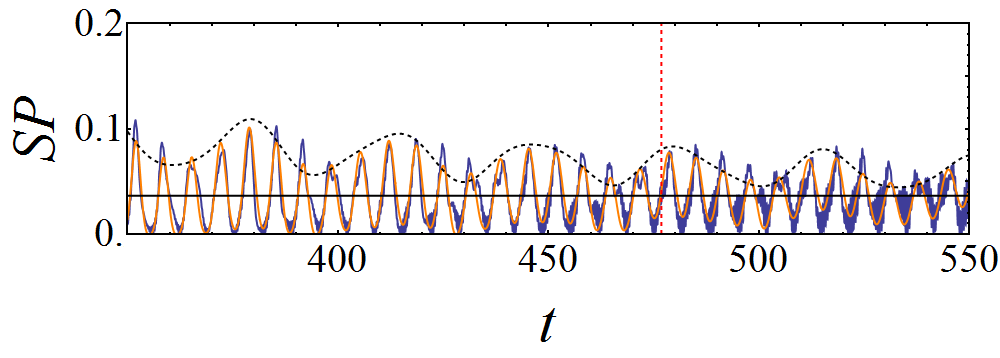}
%   $t$ &  \\
\end{tabular}
\caption{(a) Survival probability for the coherent state in figure~\ref{fig:dis} (b): numerical curve (dark blue  line), analytical approximation from  (\ref{Eq:SPwithInt}) (light orange), and  analytical modulation of the revivals (black dashed lines).   The inset shows a small time interval of the main panel to emphasize the agreement between the expression and the numerics. (b) Closer view around the decay time $t_D^{(1)}$  of the main sub-sequence of participating energy levels.  This decay time  is indicated with  vertical lines in both main panels.  The horizontal black line depicts  $IPR=0.0365$.
 }
\label{fig:SPin1}
\end{center}
\end{figure}

%%%%%

All these  contributions are now gathered into %Eq.~
(\ref{Eq:SPwithInt}), which can be compared with the numerical results. At variance with the case of a single sequence, when several sub-sequences participate in the energy eigenbasis decomposition of the coherent states, one needs to deal with several parameters to describe the evolution of the survival probability. Although they can be obtained analytically employing a semi-classical analysis, here we estimate them numerically from the exact energy spectrum. %

In figure~\ref{fig:SPin1}, we study the evolution of the survival probability for the coherent state from figure~\ref{fig:dis} (b). The parameters employed in %Eq.~
(\ref{Eq:SPwithInt}) are shown in table~\ref{parameters}. The analytical expression provides an accurate description of the numerical result from $t=0$ until the decay time $t_D^{(1)}$ of the dominant sub-sequence. The expression captures the details of the interference terms. They produce slow oscillations that modulate the fast revivals associated with the isolated sub-sequences. These slow oscillations are approximately described by the analytical expression %(\ref{Eq:SPwithInt})
 by making %the products $p\omega_1^{(i)}=0$ and
$x_i=0$  and  $p\omega_{ij}=0$ in the arguments of the Jacobi theta and  cosine functions in %Eqs.
(\ref{Eq:iscon}) and (\ref{Eq:incon}), respectively. The result is shown in both panels of figure~\ref{fig:SPin1} with dashed lines. The inset of figure~\ref{fig:SPin1} (a), which zooms in a small time interval of the main panel, reinforces the accuracy of the analytical expression.

Analogously to %the discussion associated with %Eq.~
%(\ref{Eq:tD}) and
 figure~\ref{fig:spLMG}~(b) in Sec.~\ref{Sec:LMG}, figure~\ref{fig:SPin1}~(b) makes explicit the effects of the discrete nature of the quantum spectrum, which becomes important for $t>t_D^{(1)}$. These effects are neglected by the analytical approximation, whose oscillations  beyond the decay time differ from those of the numerics.

%%%%%
\begin{figure}
\begin{tabular}{cc}
(a) & (b)\\
 \includegraphics[width=.48\textwidth]{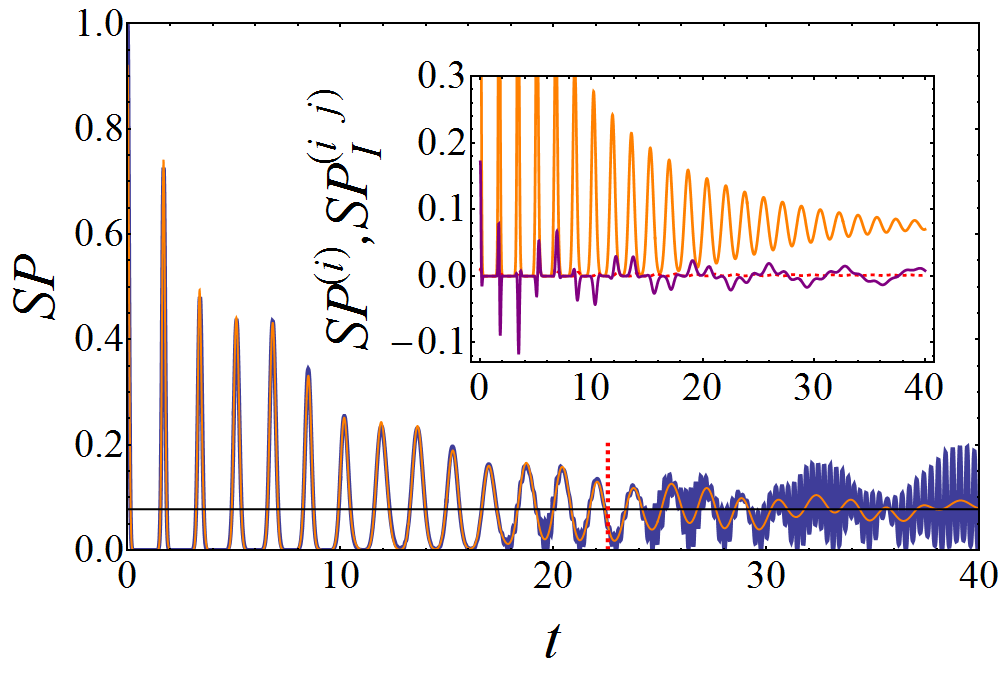}&%\rotatebox{90}{\hspace{30pt}\rotatebox{-90}{\includegraphics[width=.47\textwidth]{fig8bnew}}}
 \includegraphics[width=.48\textwidth]{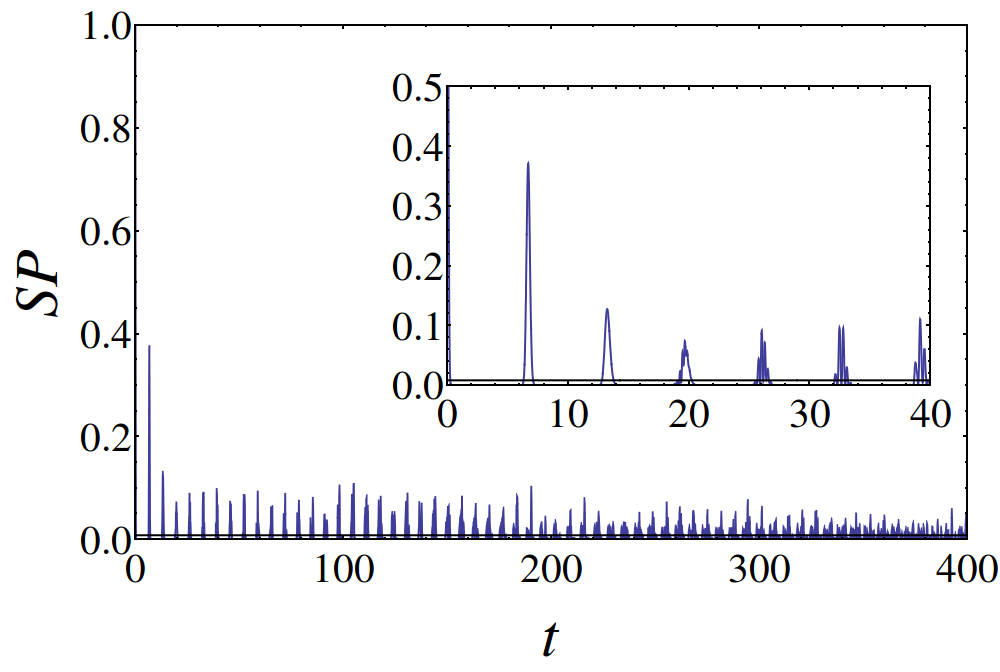}
 \end{tabular}
\caption{ (a) Survival probability for the coherent state in figure~\ref{fig:dis} (f): numerical curve (dark blue line) and analytical approximation (light orange). The decay time  $t_D^{(1)}$ of the dominant sub-sequence is indicated with a dashed vertical line. The horizontal black line depicts  $IPR=0.0777$. The inset shows $SP^{(1)}(t)$ (light orange line), $SP^{(2)}(t)$ (red dashed line), and $SP_I^{(12)}(t)$ (dark purple line). (b) Numerical result for the survival probability of the coherent state in figure~\ref{fig:dis} (d), which is located at the separatix between the region of nonlinear resonances and the region of adiabatic modes. The  inset shows the evolution in  a shorter time interval.%   Note that the vertical scale used is smaller than in panel (a) and in the previous figures. 
The  $IPR =0.00855$  is shown by a horizontal black line, but its is so close to the horizontal axis that it is difficult to distinguish it. 
}
\label{fig:SPin2}
\end{figure}
%%%%%

In figure~\ref{fig:SPin2}~(a), expression (\ref{Eq:SPwithInt}) is compared with the numerical results for the survival probability of the coherent state  from figure~\ref{fig:dis} (f), showing excellent agreement  from $t=0$ up to the decay time $t_D^{(1)}$.  
This state has two main sub-sequences, whose adjusted parameters are given in  table~\ref{parameters}. Notice that  the frequency  of the revivals is larger than in figure~\ref{fig:SPdicke} and figure~\ref{fig:SPin1}.  As discussed  in \ref{Sec:StatesDicke}, this can be qualitatively understood because the coherent state from figure~\ref{fig:dis} (f) is located in a region of the phase space corresponding to wide and  fast pseudospin excitations, in contrast with the states %in figure~\ref{fig:dis} (a) and figure~\ref{fig:dis} (b)
of figures  \ref{fig:SPdicke} and \ref{fig:SPin1},  which are dominated by the slow bosonic mode. 

On the other hand, the anharmonicities  $e_2^{(i)}$ of the state of figure~\ref{fig:SPin2}~(a) %in figure~\ref{fig:dis} (f)
 are also larger than for the states of figures \ref{fig:SPdicke} and \ref{fig:SPin1}% in Figs.~\ref{fig:dis} (a) and (b)
. This yields a decay time for the survival probability in figure~\ref{fig:SPin2}~(a) that is one order of magnitude smaller%than those in figure~\ref{fig:SPdicke} and figure~\ref{fig:SPin1}
, so fewer revivals are seen before $t_D^{(1)}$. 

Also in contrast with figure~\ref{fig:SPin1} is the almost  lack of modulation of the fast oscillations in figure~\ref{fig:SPin2}~(a). This happens, because the effect of the interference term is less pronounced, as can be confirmed in the  % than for the coherent state in figure~\ref{fig:dis}~(b). The 
inset of figure~\ref{fig:SPin2}~(a), which  shows separate curves for $SP^{(1)}(t)$, $SP^{(2)}(t)$, and $SP_I^{(12)}(t)$.% and confirms that the contribution from $SP_I^{(12)}(t)$  is small. 

\subsection{Effects of the nonlinear resonances}
\label{Sec:Resonance}

In figure \ref{fig:SPin2%fig:SPsep
}~(b), we show  the numerical result for the survival probability of the coherent state  of figure~\ref{fig:dis} (d), which  is located at the separatrix of the  nonlinear resonances region of the classical phase space, where chaos emerges. The eigenstate decomposition of this initial state is complex, with no easily identifiable structures.  This is reflected in the rapid decay of $SP(t)$, the weakness of its partial revivals (inset of the same figure),  and the fact that the $IPR$ is one order of magnitude smaller than in the previously discussed cases, where analytical approximations were applicable.

The fast decay of the survival probability signals the presence of a narrow chaotic region, which becomes larger  for higher energies. The complicated eigenstate decomposition of this initial coherent state is a quantum manifestation of classical nonlinear resonances related to %precursor of chaos reminiscent of
 the behavior of the Husimi function of the eigenstates of the %Harper 
 standard map reported in~\cite{Wisniacki2015}.

%%%%%%%%%%%%%%%%%%%%%%%%%%%% %%%%%%%%%%%%

\section{Conclusions}
\label{Sec:Concl}
We have studied, in the semi-classical limit,  the quantum dynamics of bounded systems with one- and two-degrees of freedom represented by the LMG and Dicke Hamiltonians, respectively. For this, we employed the survival probability and used coherent quantum states as initial states. Our focus was on parameters and energies for which both models have classical regular trajectories. 

Contrary to a great number of studies of the survival probability that are numerical and concentrate on some intervals of time, we obtained analytical results that cover the entire evolution, from $t=0$ to equilibration. This  allowed us to understand the onset of partial revivals, the rate of their decay, and the equilibration time. The characterization of the time scales involved in the relaxation process of isolated quantum systems is a major open question. The fact that we were able to determine the equilibration time analytically is an important contribution to the field.

 The analysis presented here is valid in general for bounded models with few degrees of freedom and can be extended to study the dynamical evolution of other observables.  For instance,  models  such as the generalized Dicke \cite{Jaako16}   and the non-linear Rabi and Dicke models \cite{Felicetti15,Penna17}  are well suited for being studied with our approach.   

%Our analysis was based on initial coherent states. 
The initial coherent states  are not only experimentally accessible, but they allow for a connection with the classical phase space. Therefore, they are a natural starting point for theoretical and experimental studies of the dynamical consequences of chaos on the survival probabilities and other physical observables.

The evolution of the survival probability depends on the energy distribution of the initial state.
 For one-degree of freedom systems and for  two-degrees of freedom systems  when only one of the two degrees of freedom is excited,   the spectrum of the energy states contributing to the initial state is quasi-harmonic with  Gaussian weights.  In this case, an expression for the survival probability in terms of the Jacobi theta function was derived and shown to be in excellent agreement with the  numerical results.  The expression describes the periodic partial revivals of
the initial state and the slow equilibration. We also found that the equilibration time, given by the inverse of the anharmonicity parameter, diverges in the classical limit.

In  more complex situations where the two-degrees of freedom are excited, the dynamics is determined by several interference terms that result in the modulation of the revivals. For the regular regime, we were still able to derive an analytical expression. As the system approaches  chaotic regions,  the distribution of the components of the initial state loses a simple recognizable structure, resulting in very short equilibration times.

An interesting future direction is to connect the results of this work with those of %Ref.~
\cite{Altland2012NJP}, where the temporal  evolution of initial coherent states under the Dicke model was also studied. 
We conjecture that the analytical formula given here is related to the classical drift term in %Ref.~
\cite{Altland2012NJP}, while the fluctuations observed at times longer than the equilibration time are related to a  diffusive quantum term.   

\section*{Acknowledgments}
%\ack
We acknowledge financial support from Mexican CONACyT project CB2015-01/255702, DGAPA- UNAM project IN109417 and RedTC. MABM is a post-doctoral fellow of CONACyT. S.L-H. acknowledges financial support from the CONACyT fellowship program for sabbatical leaves. LFS is supported by the NSF grant No. DMR-1603418.
%\end{acknowledgments}

%%%%%%%%%%%%%%%%%%%%%%%%%%%%%%%%%%%%
\appendix

\section*{Appendices}
\setcounter{section}{0}

\section{Standard deviations of the Hamiltonians in coherent states}
\label{App2}

The standard deviations of the LMG and Dicke Hamiltonians in coherent states  were calculated in \cite{Schliemann2015}. Here, we simply correct  some   misprints found in that reference.%  of %Ref.~
%, where the energy fluctuations of the LMG and Dicke models were calculated to illustrate a general property of any Hamiltonian in the coherent state basis.

According to equation~(69) of %Ref.~
\cite{Schliemann2015},  and after redefining parameters to be consistent with our parametrization for the LMG model (we have also  corrected  two misprints in the third and fourth line of Eq.(70) in   %Ref.~
\cite{Schliemann2015}), the standard deviation of the Hamiltonian in a coherent state is 
\begin{equation}
\sigma_{LMG}^2=\langle z | H_{LMG}^2|z\rangle-\langle z | H_{LMG}|z\rangle^2=\Omega_1+\Omega_2
\label{eq-app2:sigLMG}
\end{equation} 
where  (defining $\cos\theta=j_z$%/J$
)
$$
\Omega_1= \frac{J}{2}\left[-2\gamma_x\cos\theta\sin^2\theta\cos^2\phi-2\gamma_y \cos\theta\sin^2\theta\sin^2\phi\right.
$$
$$
+\gamma_x^2(\sin^4
\theta\cos^2\phi\sin^2\phi+\cos^2\theta\sin^2\theta\cos^2\phi)
+\gamma_y^2(\sin^4\theta\cos^2\phi\sin^2\phi+\cos^2\theta\sin^2\theta\sin^2\phi)
$$
$$
\left. +\sin^2\theta-2\gamma_x\gamma_y \sin^4\theta\cos^2\phi\sin^2\phi\right],
$$ 
is of order $J$.  In the  limit $J>>1$, this is the dominant term of $\sigma_{LMG}$, since   
$$
\Omega_2=\frac{1}{8}\left(1-\frac{1}{2J}\right)\left[-4\gamma_x \gamma_y \cos^2\theta
+(\gamma_x(1-\sin^2\theta\cos^2\phi)+\gamma_y(1-\sin^2\theta\sin^2\phi))^2\right]
$$ 
is of order $J^0$ and $J^{-1}$.
From the previous expressions, it is clear that in the limit $J\rightarrow\infty$, the uncertainty of $H_{LMG}$ scales as
\begin{equation}
%\sigma_{LMG}\propto \sqrt{J}.
\sigma_{LMG}\approx f_{\sigma} \sqrt{J}.
\label{sigma}
\end{equation}
%%%%%%%%%%%%%%%%%%%%
Similarly,  the uncertainty of the Dicke Hamiltonian in coherent states is depicted by equation~(85) of %Ref.~
\cite{Schliemann2015}. The result in our parametrization (and after correcting a misprint in the third line of equation~(86) in %Ref.~
\cite{Schliemann2015}) is
$$
\sigma_{D}^2=\langle z\alpha | H_{D}^2|z\alpha\rangle-\langle z\alpha | H_{D}|z\alpha\rangle^2=\Omega_1+\Omega_2,
$$ 
where, again, $\Omega_1$ is linear in $J$, 
$$
\Omega_1=J\left\{ \frac{\omega^2}{2}(q^2+p^2)+\frac{\omega_o^2}{2}\sin^2\theta
+2\gamma^2\left[ (\sin^2\theta\sin^2\phi+\cos^2\theta)q^2+\sin^2\theta\cos^2\phi\right]\right.
$$
$$
+2\gamma q  ( \omega\cos\phi -\omega_o \cos\theta\cos\phi)\sin\theta \text{\LARGE{$ \} $}},    
$$
and gives the leading contribution in the limit $J>>1$, because  $\Omega_2$ if of order $J^0$ 
$$
\Omega_2=\gamma^2 (\sin^2\theta\sin^2\phi+\cos^2\theta).
$$
%where we have used the scaled variables for the bosonic coherent state. 
%$q_e=q/\sqrt{J}$ and $p_e=p/\sqrt{J}$, and $\cos\theta=j_z/J$.  
Therefore,  the uncertainty of the Dicke Hamiltonian in a coherent state also  scales as
 $
 \sigma_{D}\propto \sqrt{J}.
%\sigma_{D}\approx f_{\sigma} \sqrt{J}.
 $
\section{Semi-classical expansion for the spectrum}
\label{App6}

From the  Bohr-Sommerfeld quantization rule %applied to the pseudospin canonical variables, we have
 using scaled variables %$\tilde{j}_z=j_z/J$,
  ($h=H/J$ and  $\epsilon_n=E_n/J$) 
 $$
I(\epsilon_{n})=\displaystyle{\oint_{h(j_z,\phi)=\epsilon_n}} \!\!\!\!\!\!\!\!\!\!\!\!\!\!\!\!\!\!j_z d\phi= \frac{2\pi}{J}  \left(n+\frac{1}{2}\right),
$$
we obtain, for two quantized energy levels,  $\epsilon_n$ and $\epsilon_{n'}$, 
\begin{equation}
I(\epsilon_n)-I(\epsilon_{n'})=\frac{2\pi}{J} k \ \ \ \ \  \ (k=n-n').
\label{eq:anac}
\end{equation}
On the other hand,  in the classical limit $J\rightarrow\infty$, the  energy states contributing significantly to a given initial coherent state, are located in an scaled interval whose width (let us say $[(\bar{E}-3.5\sigma)/J,(\bar{E}+3.5\sigma)/J]$) goes  to zero. Consequently, the action variables associated to two quantized energies in this interval  can be approximated by a Taylor expansion 
\begin{equation}
I(\epsilon_n)- I(\epsilon_{n'})\approx \frac{2\pi}{\omega_{cl}}(\epsilon_n-\epsilon_{n'})-\frac{\pi }{\omega_{cl}^2}\frac{d\omega_{cl}}{d\epsilon}(\epsilon_n-\epsilon_{n'})^2,
\label{eq:tay}
\end{equation}
where we have used the classical relation $I'(\epsilon_{n'})=2\pi/\omega_{cl}$. 
By  equating (\ref{eq:anac}) and (\ref{eq:tay}), solving the quadratic equation for $\epsilon_n$ and expanding in powers of $1/J$, we obtain for the non-scaled energies% ($E_n=J \epsilon_n$)
$$
E_n=E_{n'}+\omega_{cl}k+\frac{\omega_{cl}}{2J}\frac{d\omega_{cl}}{d\epsilon}k^2+\mathcal{O}(J^{-2}).
$$
This relation justifies the expansion (\ref{principalspectrum}) and allows to obtain semi-classical estimates for its parameters
$$
e_1\approx\omega_{cl}\approx E_{n+1}-E_n + \mathcal{O}(J^{-1})  {\hbox{\ \ \ \ \ \ and \ \ \  \ \ \ }} e_2\approx\frac{\omega_{cl}}{2J}\frac{d\omega_{cl}}{d\epsilon}. 
$$
Since  $\omega_{cl} d\omega_{cl}/d\epsilon$ is a finite value in the  limit $J\rightarrow\infty$, the anharmonicity parameter $e_2$ goes to zero.  The vanishing of the anharmonicity  is a subtle reflect of the classical limit. In this limit, the classical (scaled) energy width of the coherent state become infinitely small,  simultaneously the number of energy states participating in the coherent state goes to infinity [see (\ref{IPR})]. In this way,  we have an infinitely narrow classical energy interval with an infinite number of quantum energy levels inside.  Since only a single classical frequency ($\omega_{cl}$) is associated to an infinitely small classical energy interval (in effective one degree-of-freedom  systems), the quantum energy levels inside the interval must be equally spaced  ($E_{n+1}-E_n\approx \omega_{cl}$), consequently  $e_2$ must be zero.

%%%%%%%%%%%%%%%%%%%%%%%%%%%%%%%%%%%%%%%555
%%%%%%%%%%%%%%%%%%%%%%%%%%%%%%%%%%%%%%

\section{Frequencies distribution, product $\pmb{ g_{k+p} g_k }$ 
%$p$-th component frequency distribution
}
\label{App5}
In this section we show that the product $ g_{k+p} g_k$  can be approximated by a single Gaussian for the frequencies $\omega_k^{(p)}=E_{k+p}-E_k$. %The derivation  distributions of the frequencies for the $p$-th component $\omega_k^{(p)}=E_{k+p}-E_k$ in the $SP_p(t)$ terms of the Survival Probability are very well reproduced by Gaussian functions. The derivation
 Based on  (\ref{components}) and (\ref{principalspectrum})
 % 
% on the observed Gaussian  for the components
%$$
%|c_k|^2\approx g_k\equiv A  e^{-\frac{(E_k-\bar{E})^2}{2\sigma^2}}, 
%$$
%and the assumption for the spectrum 
%\begin{equation}
%E_k=e_o+e_1 k+e_2 k^2.
%\label{eq:app51}
%\end{equation}
%
%We start with the product of Gaussian functions% for the components separated an index distance $p$  
\begin{eqnarray}
%%g_k g_{k+p}&= A^2 \exp\left[-\frac{(E_k-\bar{E})^2+(E_{k}-\bar{E}+\omega_k^{(p)})^2}{2\sigma^2}\right] \nonumber\\
%$$
%Our objective is to express this product entirely in terms of $\omega_k^{(p)}$. To this end, we rewrite it as
%\begin{equation}
%g_k g_{k+p}
%%&= A^2 \exp\left[-\frac{\left(\omega_k^{(p)}\right)^2}{4\sigma^2}\right] 
%%\exp\left\{-\frac{1}{\sigma^2}\left[E_k-(\bar{E}-\omega_k^{(p)}/2)\right]^2\right\}\\
&g_k g_{k+p}= A^2 \exp\left[-\frac{\left(\omega_k^{(p)}\right)^2}{4\sigma^2}\right]\exp\left[-\frac{F\left(\omega_k^{(p)}\right)}{16 \sigma^2 e_2^2 p^4} \right], 
\label{eq:app52}
\end{eqnarray}
%To express $E_k$ as a function of $\omega_k^{(p)}$, we use %Eq.~
%(\ref{eq:app51}) and obtain
%$$
%\omega_k^{(p)}=E_{k+p}-E_k=p(e_1+p e_2)+2pe_2 k,
%$$
%from where
%$$
%k=\frac{\omega_k^{(p)}-p(e_1+pe_2)}{2 pe_2},
%$$ 
%which plugged into %Eq.~
%(\ref{eq:app51}) 
%In the last line
where we have we used (\ref{omegap}) to express $E_k$ as a quadratic function of $\omega_k^{(p)}$,  
$$
E_k=e_o + \frac{(\omega_k^{(p)}-e_2 p^2)^2- e^{2}_1 p^2}{4 e_2 p^2}, 
$$
%By substituting this expression in (\ref{eq:app52}), the product of Gaussians reads
%$$
%g_k g_{k+p}= A^2 \exp\left[-\frac{\left(\omega_k^{(p)}\right)^2}{4\sigma^2}\right]\exp\left[-\frac{F\left(\omega_k^{(p)}\right)}{16 \sigma^2 e_2^2 p^4} \right],
%$$
%with 
and defined
$$F\left(\omega_k^{(p)}\right)=\left\{\left(\omega_k^{(p)}\right)^2+p^2\left(4e_2(e_o-\bar{E})+e_2^2 p^2-e_1^2\right)\right\}^2. $$   
Since the function $F$ %\left(\omega_k^{(p)}\right)$ 
is inside the argument of the exponential, the frequencies $\omega_k^{(p)}$ far from its minimum, $\omega_{p}$, % of $F\left(\omega_k^{(p)}\right)$ 
are highly suppressed. %For our purpose is enough to consider up to quadratic terms around the minimum $\omega_{p}$
%$$
%g_k g_{k+p}\approx  A^2 \exp\left[-\frac{\left(\omega_k^{(p)}\right)^2}{4\sigma^2}\right]
%\exp\left[-\frac{F\left(\omega_p\right)+ F''(\omega_p)(\omega_k^{(p)}-\omega_p)^2/2}{16  e_2^2 p^4 \sigma^2} \right].
%$$
%By calculating  the minimum and evaluating $F$ and $F''$ at it, we obtain
Expanding up to second order around $\omega_{p}$ and using %that
\begin{equation}
\omega_{p}=p\sqrt{e_1^2+4 e_2(\bar{E}-e_0)-e_2^2 p^2}, {\hbox {\ \ \ \ }} F\left(\omega_{p}\right)=0 {\hbox{\ \ and \ \ }}F''\left(\omega_{p}\right)/2=4 \omega_{p}^2,
\label{eq:app53}
\end{equation}
%$$
%F\left(\omega_{p}\right)=0,
%$$ 
%$$
%F''\left(\omega_{p}\right)/2=4 \omega_{p}^2,
%$$
%With these results, we obtain the following distribution for the 
the distribution of
frequencies of the $p$-th component,
can be written as a product of two Gaussians.
%$$
%g_k g_{k+p}\approx  A^2 \exp\left[-\frac{\left(\omega_k^{(p)}\right)^2}{4\sigma^2}\right]\exp\left[-\frac{\omega_p^2(\omega_k^{(p)}-\omega_p)^2}{4 e_2^2 p^4 \sigma^2} \right],
%$$
%which is  a product of two Gaussians, one centered in $0$ and having width $\sqrt{2}\sigma\propto\sqrt{J}$  and a second much narrower  with centroid in $\omega_p\propto J^0$ and width 
%\begin{equation}
%\sigma_p= \sqrt{2}\,|e_2|\,p^2 \frac{\sigma}{\omega_{p}}, 
%\label{eq:app54}
%\end{equation} 
%whose dependence on $J$ can be shown to be  $\sigma_p\propto J^{-1/2}$ (see  \ref{App2} for the scaling of $\sigma$ and \ref{App6} for the scaling of $e_2$). 
For large $J$,  since the width of the second Gaussian is very narrow, 
%the product of Gaussians is highly suppressed, except in the region around $\omega_{k}^{(p)}\approx \omega_p$. Therefore the product of Gaussians 
it can be reduced to 
%\begin{equation}
%g_k g_{k+p}\approx  A^2 \exp\left[-\frac{\omega_p^2}{4\sigma^2}\right]\exp\left[-\frac{(\omega_k^{(p)}-\omega_p)^2}{2\sigma_p^2} \right], 
%\end{equation}
%where the contribution from the first Gaussian is its value at the centroid of the second Gaussian.
%The final result is a single Gaussian function for every $p$-th component, 
$$
g_k g_{k+p}\approx  A_p\exp\left[-\frac{(\omega_k^{(p)}-\omega_p)^2}{2\sigma_p^2} \right],
$$
 with amplitude 
$
\frac{A_p}{A^{2}}= \exp\left[-\frac{\omega_{p}^2}{4\sigma^2}\right]
$
% centroid $\omega_p$ (\ref{eq:app53})  
and width $\sigma_p = \sqrt{2} |e_2| p^2 \sigma/\omega_{p}$.% (\ref{eq:app54}).
 
 Finally, since $|e_2|<< |e_1|$, at leading order in $e_2$, the centroid, width, and amplitude of the $p$-th  distribution 
are given simply in terms of the values for $p=1$
%$$
%\omega_{p}\approx p \sqrt{e_1^2+4 e_2(\bar{E}-e_0)}\approx p \,\omega_{1} ,
%$$ 
%\begin{equation}
%\sigma_p\approx\sqrt{2}\,|e_2| \, p^2\frac{\sigma}{p\,\omega_{1}}=\sqrt{2}\, |e_2| \, p\frac{\sigma}{\omega_{1}}\approx p \,\sigma_{1} ,
%\label{eq:app56}
%\end{equation}
%$$
%\frac{A_p}{A^2}\approx \exp\left(-\frac{\omega_{p}^2}{4\sigma^2}\right)\approx\exp\left(-\frac{\omega_{1}^2}{4\sigma^2} p^2\right)\approx
%(A_1)^{p^2}.
%\left(\frac{A_1}{A^2}\right)^{p^2},
%$$
%with 
%$$
%\omega_{1}=\sqrt{e_1^2+4e_2 (\bar{E}-e_0)-e_{2}^{2}}\approx\sqrt{e_1^2+ 4 e_2 (\bar{E}-e_0)},
%$$
%$$
%\sigma_{1}=\sqrt{2}\, |e_2| \, \frac{\sigma}{\omega_{1}},
%$$
%and
%$$
%A_1=A^2 \exp\left(-\frac{\omega_{1}^2}{4\sigma^2} \right).
%$$
%
%
%%$$
%%\omega_{p} \approx p \,\omega_{1} ,
%5$$ 
\begin{equation}
\omega_{p} \approx p \,\omega_{1} , \ \ \ \sigma_p \approx p \,\sigma_{1} \ \ {\hbox{ and}} \  \ \ \  \frac{A_p}{A^2}\approx
\left(\frac{A_1}{A^2}\right)^{p^2}
\label{eq:app56}
\end{equation}
%%$$
%%\frac{A_p}{A^2}\approx
%%\left(\frac{A_1}{A^2}\right)^{p^2},
%%$$
with 
$$
\omega_{1}\approx \sqrt{e_1^2+ 4 e_2 (\bar{E}-e_0)}
%%\sqrt{e_1^2+ 4 e_2 (\bar{E}-e_0)}
, \ \  \ \sigma_{1}=\sqrt{2}\, |e_2| \, \frac{\sigma}{\omega_{1}}, \ \ {\hbox{and}} \ \ \ \frac{A_1}{A^2}= \exp\left(-\frac{\omega_{1}^2}{4\sigma^2} \right).
$$
%%$$
%%$$
%%\sigma_{1}=\sqrt{2}\, |e_2| \, \frac{\sigma}{\omega_{1}},
%%$$
%%and
%%$$
%%A_1=A^2 \exp\left(-\frac{\omega_{1}^2}{4\sigma^2} \right).
%%$$

\section{Approximation of the sum by an integral for the $p$-th component of the survival probability}
\label{App7}

Here, we approximate the sum appearing in the the $p$-th component of the $SP$ %Equation 
(\ref{SPppp}) %(Eq.~(19) in \cite{mainArticle})
 by an integral. Using our assumption for the principal spectrum %[Equation   
 (\ref{principalspectrum}),  we write  the differences between consecutive frequencies as
 $\Delta\omega_k^{(p)}=\omega_{k+1}^{(p)}-\omega_{k}^{(p)}%$ as
 =2\, p\, e_2$
%$$
 %\Delta\omega_k^{(p)}=2pe_2,
%$$
and arrive at 
%\begin{align}
%SP_p(t)= \dfrac{\omega_1^2 }{\pi \sigma^2}   \exp\left(-\frac{\omega_1^{2}}{4\sigma^{2}}p^{2} \right)  
%\displaystyle\sum_k 
%\exp\left[-\frac{\left(\omega_k^{(p)}-p \omega_1\right)^2}{2 p^2 \sigma_1^2}\right] \cos(\omega_k^{(p)}  t)
%\frac{ \Delta\omega_k^{(p)}  }{2pe_2}
 %\label{eq:sumtoint} \\
 %\approx \dfrac{\omega_1^2 }{\pi \sigma^2 2p |e_{2}|}  \exp\left(-\frac{\omega_1^{2}}{4\sigma^{2}}p^{2}\right) \int \exp\left(-\frac{(\omega-p\,\omega_1)^2}{2 p^2 \sigma_1^2}\right) \cos(\omega  t)d\omega \nonumber.
%\end{align}
%The integral can be calculated straightforwardly
%\begin{align}
%\int \exp\left(-\frac{(\omega-p\,\omega_1)^2}{2 p^2 \sigma_1^2}\right) \cos(\omega  t)d\omega=
%\mathbf{Re}\int  \exp\left(-\frac{(\omega-p\,\omega_1)^2}{2 p^2 \sigma_1^2}\right) e^{i \omega  t}d\omega \nonumber\\
%=\mathbf{Re}\left[ e^{i p \omega_1 t}\int  \exp\left(-\frac{(\omega-p\,\omega_1)^2}{2 p^2 \sigma_1^2}\right) e^{i (\omega-p\,\omega_1)  t}d\omega\right]=
%\mathbf{Re}\left[ e^{i p \omega_1 t}\sqrt{2\pi} p \sigma_1 e^{-\frac{(p\sigma_1 t)^2}{2}}\right]\nonumber\\
% =\sqrt{2\pi} p \sigma_1 \exp\left[-\frac{(p\sigma_1 t)^2}{2}\right]\cos(p\,\omega_1  t). \nonumber
%\end{align}

\begin{align}
\displaystyle\sum_k 
\exp\left[-\frac{\left(\omega_k^{(p)}-p \omega_1\right)^2}{2 p^2 \sigma_1^2}\right] \cos(\omega_k^{(p)}  t)
%\frac{ \Delta\omega_k^{(p)}  }{2pe_2}
 %\label{eq:sumtoint} \\
 \approx \dfrac{1 }{ 2 \,p |e_{2}|} 
 \int \exp\left(-\frac{(\omega-p\,\omega_1)^2}{2 p^2 \sigma_1^2}\right) \cos(\omega  t)d\omega \nonumber
\end{align}
%The integral can be calculated straightforwardly
\begin{align}
%\int \exp\left(-\frac{(\omega-p\,\omega_1)^2}{2 p^2 \sigma_1^2}\right) \cos(\omega  t)d\omega=
%\mathbf{Re}\int  \exp\left(-\frac{(\omega-p\,\omega_1)^2}{2 p^2 \sigma_1^2}\right) e^{i \omega  t}d\omega \nonumber\\
%=\mathbf{Re}\left[ e^{i p \omega_1 t}\int  \exp\left(-\frac{(\omega-p\,\omega_1)^2}{2 p^2 \sigma_1^2}\right) e^{i (\omega-p\,\omega_1)  t}d\omega\right]=
 = \dfrac{1 }{ 2 \,p |e_{2}|}  \mathbf{Re}\left[ e^{i p \omega_1 t}\sqrt{2\pi} p \sigma_1 e^{-\frac{(p\sigma_1 t)^2}{2}}\right]
 = \dfrac{\sqrt{\pi} \sigma_1 }{ \sqrt{2}  |e_{2}|}
 \exp\left[-\frac{(p\sigma_1 t)^2}{2}\right]\cos(p\,\omega_1  t). \nonumber
\end{align}
With the result above, the expression for $\sigma_1$ (\ref{eq:aproxDis}) %in \cite{mainArticle}) 
and the considerations for equation~(\ref{iden}), we obtain the following expression for the $p$-th component of the $SP$
\begin{equation}
SP_p(t)=\frac{\omega_1}{\sigma\sqrt{\pi}} \exp\left[-p^{2}\left(\frac{\omega_{1}^2}{4\sigma^2} +\frac{t^2}{t_D^2}\right)\right]\cos(p\,\omega_1  t),
\label{eq:app61}
\end{equation}
where we have defined the decay time %of the $p=1$ component as
$
t_D\equiv\frac{\omega_1}{ \sigma |e_2|  }.
$

\section{Parameter dependence on the coordinates of the initial coherent state in the LMG model.}
\label{AppPar}
\begin{figure}
\begin{tabular}{rcrc}
\rotatebox{90}{\ \ \ \ \ \ \  \ \ \ \ \  \ \ \ \  \ \ \ $(1+j_z)\sin\phi$} &\includegraphics[width=0.42\textwidth]{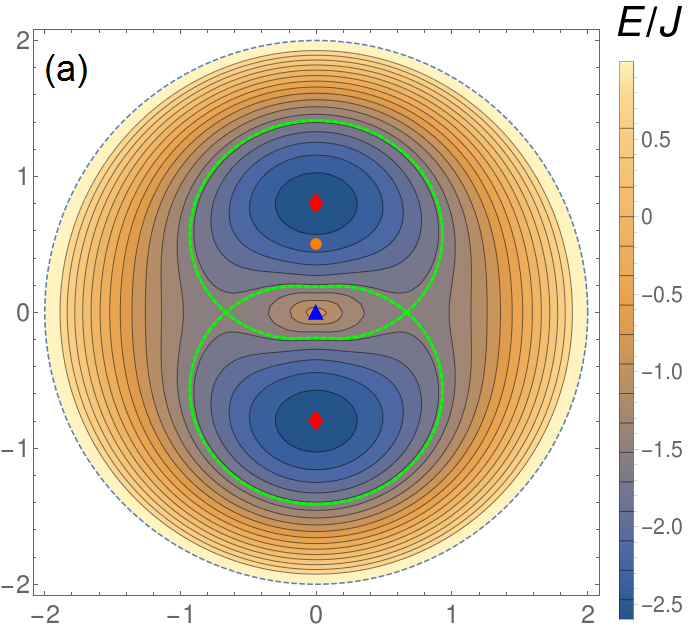} & \rotatebox{90}{\ \ \ \ \ \ \  \ \ \ \ \  \ \ \ \  \ \ \ $|e_2|$} &\rotatebox{90}{\hspace{10pt}\rotatebox{-90}{\includegraphics[width=.40\textwidth]{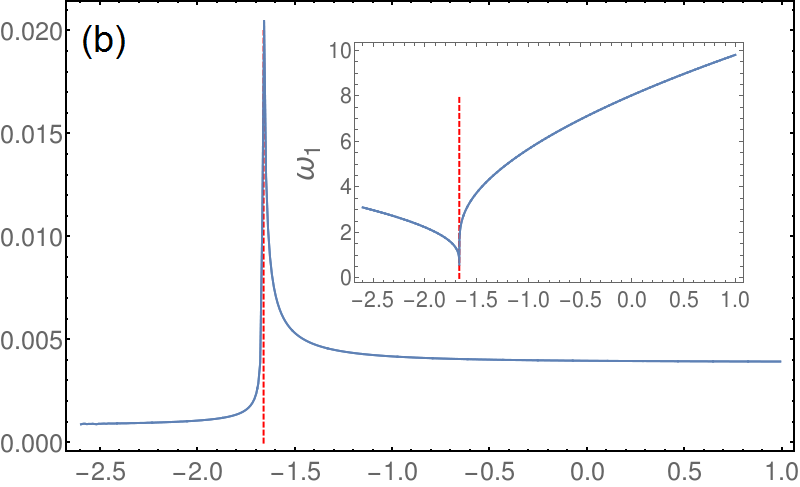}}}\\%\includegraphics[width=0.40\textwidth]{2igAb}\\
 &$(1+j_z)\cos\phi$ &  & $E/J$ \\
\rotatebox{90}{\ \ \ \ \ \ \  \ \ \ \ \  \ \ \ \  \ \ \ $(1+j_z)\sin\phi$} &\includegraphics[width=0.42\textwidth]{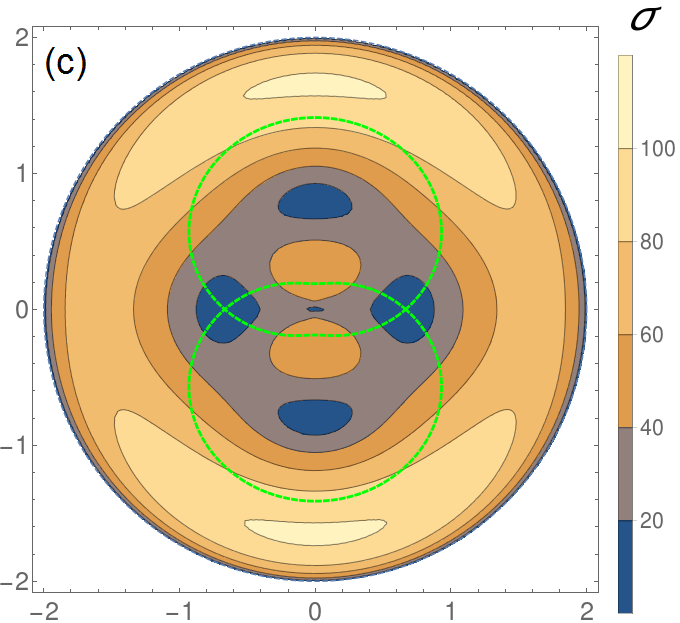} & & \includegraphics[width=0.42\textwidth]{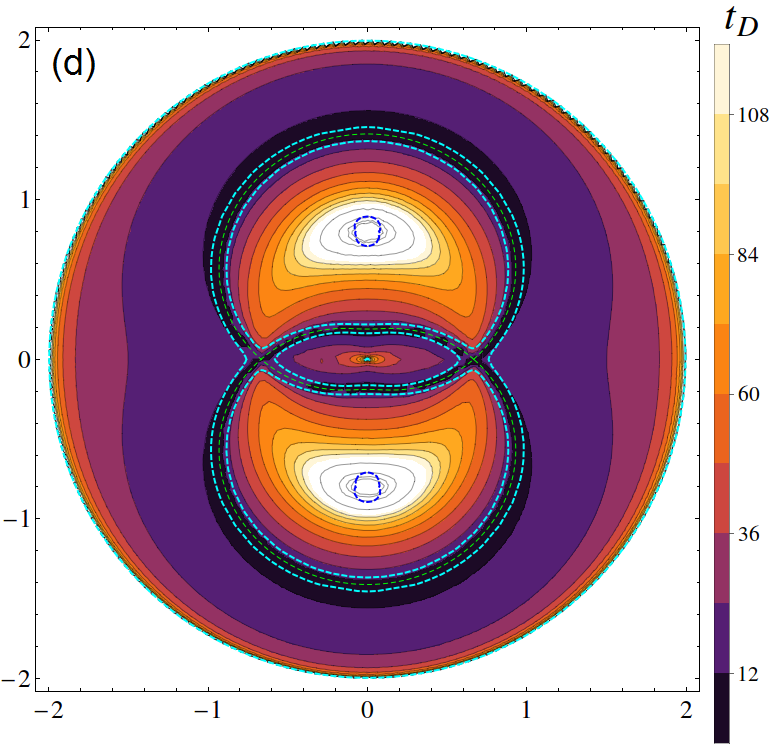}\\
 & $(1+j_z)\cos\phi$ & & $(1+j_z)\cos\phi$\\
\end{tabular}  
\caption{ Scaled energy (a),  energy width (c) and time decay (d) dependence on the parameter of the initial coherent state for the LMG model. (b) Anharmonicity parameter and  classical frequency $\omega_{1}$ (inset) as a function of scaled energy. See text for details.}
\label{fig:Nueva}
\end{figure}
Using the semi-classical formulae for $\omega_1$ (\ref{limclassom}) and $e_2$ (\ref{e2sc}), and the analytical expression for the energy standard deviation $\sigma$ (\ref{eq-app2:sigLMG}), we study  their  dependence and that of the decay time  $t_D=\omega_1/(\sigma |e_2|)$ (\ref{Eq:tD}) on the parameters of the initial coherent states for the LMG model with couplings as in section \ref{Sec:LMG} of the main text, $\gamma_x=-3, \gamma_y=-5$, and $J=2000$. A similar analysis can be performed for any other coupling set. The coherent parameter space defines  the surface of the so-called Bloch sphere. For simplicity, this surface is represented in a 2D circle using coordinates $(1+j_z)\cos\phi$ and $(1+j_z)\sin\phi$. In this representation, the south pole is located in the center of the circle whereas the north pole is deformed to the outer circle of radius $2$.    $\omega_1$ and $e_2$ are obtained by using the method described in \cite{Engelhardt2015}. 

The results are shown in figure \ref{fig:Nueva}. In panel (a),  level curves  of the scaled energy  are shown.  Red diamonds indicate the ground-state configurations with $E/J=-2.6$. Dashed green line indicates the level curve of the ESQPT critical energy at $E_{ESQPT}/J=-1.6667$. The central blue triangle is a local maximum critical point at energy $E/J=-1$.  The orange dot indicates the coordinate of the representative coherent-sate discussed in section \ref{Sec:LMG} of the  main text.    The parameters $e_2$  and $\omega_1$   depend only on energy, their  dependences are shown, respectively,  in panel (b)  and inset. The vertical dashed lines indicate the ESQPT critical energy, where $e_2$ diverges and $\omega_1=0$.  Contrary to the latter parameters, the energy width $\sigma$ depends on the localization of  the  coherent states   in the Bloch sphere, this dependence is shown in  panel (c).  The decay time of the $SP$ ($t_D$) as  a function of the initial coherent state is shown in panel (d). In this panel, blue and cyan  lines delimit the small regions around the critical energy configurations (ground-state  and ESQPT respectively)  where our approach is not applicable for $J=2000$ (we use the criterion $E\pm3.5 \sigma=E_{cr}$ to define these lines). These regions become narrower as $J$ approaches the classical ($J\rightarrow\infty$) limit. 
Observe that the decay time increases unboundedly  for states approaching the ground state configuration or  the local maximum in the center, whereas for states close to the ESQPT the decay time  goes to zero.  This latter  result is in accord with those of \cite{Fernandez2011}. 
 
A similar  analysis of the $SP$ parameters for the Dicke model, is a challenging task, not only because the dimension of the energy surfaces is three, but also because the increase in the numbers of Gaussian sub-sequences makes it hard to provide a complete analysis. However,  the different study cases used in the main text are representative of the whole cases that can be found in the regular regime of the Dicke model.
%%%%%%%%%%%%%%%%%%%%
%%%%%%%%%%%%%%%

%%%%%%%%%%%%%%%%%%%%
\section{Classical limit of the survival probability and power law decay of its revivals} 
\label{App6bis} 
In this section we derive analytically the classical limit of the survival probability,  as well as the power law decay of its revivals.
 
The decay of the oscillations in the survival probability is given by 
%We start with
 the expression (\ref{theta}) with the first argument $x=0$. Taking the limit  
%We begin from $SP^{Decay}(t)$, that is, %Eq.~
%(\ref{SPsuman}) %in \cite{mainArticle} 
%without the cosine function, which reads
%$$
%SP^{Decay}(t) \! =\! \frac{\omega_1}{2\sigma\sqrt{\pi}}
%\left\{\!\! 1+2 \! \displaystyle\sum_{p=1} \left[\! \exp\left(-\frac{\omega_1^2}{4\sigma^2}-  \frac{t^2}{t_D^2}\right) \right]^{p^2} \right\}.
%$$ 
%In terms of the Jacobi theta  function, it is
%$$
%SP^{Decay}(t) \! =\! \frac{\omega_1}{2\sigma\sqrt{\pi}}\Theta_3(0,y),
%$$
%with $y=e^{-\frac{1}{4}\left(\frac{\omega_1}{\sigma}\right)^2} e^{-\left(\frac{t}{t_D}\right)^2}$. 
%
%By considering the asympotic (
$J>>1$ %) behaviors of 
and using 
$\omega_1\approx \omega_{cl}$, $\sigma\approx f_\sigma\sqrt{J}$, $e_2\approx f_e/J$ and $t_D=\omega_1/(\sigma |e_2|)$,
 the SP decay reads
$$
SP^{Decay}(t) \! \approx\!\frac{\omega_{cl}}{2 f_\sigma\sqrt{\pi}\sqrt{J}}\Theta_3\left(0,e^{-\frac{1}{J}\left[\frac{\omega_{cl}^2}{4 f_\sigma^2}+\frac{f_\sigma^2 f_e^2}{\omega_{cl}^2}t^2\right] }\right).
$$
%where we have used $t_D=\omega_1/(\sigma |e_2|)$. 
%Now, from the properties of the Jacobi theta function it can be shown that 
As $\lim_{J\rightarrow\infty}\Theta_3(0,e^{-b/J})/\sqrt{J}=\sqrt{\pi/b}$,%. By using this result it is straightforward to show that 
\begin{equation}
\lim_{J\rightarrow\infty}SP^{Decay}(t)= \frac{1}{\sqrt{1+\left(\frac{2 f_\sigma^2 f_e t}{\omega_{cl}^2} \right)^2}}
\approx \frac{\omega_{cl}^2}{2 \sigma^2 |e_2|}\frac{1}{t} ~~~(t>>1),
\label{eq:SPDecAsym}
\end{equation}%
%At  times of the order of $t_D>>1$, the previous expression can be  approximated as 
%$$
%\lim_{J\rightarrow\infty}SP^{Decay}(t) \approx  \frac{\omega_{cl}^2}{2 f_\sigma^2 |f_e|}{\frac{1}{t}}=\frac{\omega_{cl}^2}{2 \sigma^2 |e_2|}\frac{1}{t},
%$$
which explains the power-law decay observed in the survival probability at times $t\sim t_D$. 
%For the particular case   of  figure\ref{fig:spLMG}, % in \cite{mainArticle} % For that particular case  %of the parameters and initial state for that case 
% the factor  $\frac{\omega_{cl}^2}{2 \sigma^2 |e_2|}\approx 2.512$ and the decay is  $SP^{Decay}(t)\approx 2.512/t$,  which compares very well with the numerical fit to the decay ($2.506/t$).

%For the entire expression of the $SP(t)$ %Eq.~
%(\ref{SPsuman})%in \cite{mainArticle}
 The partial revivals occurring at integer multiples of the classical period $\tau=2\pi/\omega_{cl}$ become more and more narrower as $J$ increases, turning into  Kronecker deltas in the limit $J \rightarrow\infty$. The heights ($f_n$) of the widthless  revivals are given by % Eq.
  (\ref{eq:SPDecAsym}) evaluated  in $t=n\tau= 2\pi n/\omega_{cl}$, thus  the  $SP$  in the  limit $J\rightarrow\infty$ is
$$
\lim_{J\rightarrow\infty}SP(t)=\sum_{n\in\mathbb{Z}}\delta_{t,n\tau}f_n,  
{\hbox{\ \ \ with \ \ \ }}
f_n
= \frac{1}{\sqrt{1+\left(\frac{4 \pi f_\sigma^2 f_e}{\omega_{cl}^3}\right)^2 n^2}}.
$$
%where in the rightmost term  we have used the expression for $f_e$ in %Equation
% (\ref{eq:edsc}).%  in  \ref{App6}.% of this supplemental material.

%%%%%%%%%%%%%%%%%%%%%%%%%
%%%%%%%%%%%%%%%%%%%%%%%%%

\section{Interference terms}
\label{App8}
%The survival probability for the case where we have several sub-sequences (indexed by $i,j$) can be written as
%$$
%SP(t)=\left|\sum_{ik} |c_k^{(i)}|^2 e^{i E_k^{(i)}t}\right|^2=
%\sum_{ij}\sum_{kk'}|c_k^{(i)}|^2|c_{k'}^{(j)}|^2 e^{i\left(E_k^{(i)}-E_{k'}^{(j)}\right)t}
%$$
%$$
%=\sum_{i}\sum_{kk'} |c_k^{(i)}|^2|c_{k'}^{(i)}|^2 e^{i\left(E_{k}^{(i)}-E_{k'}^{(i)}\right)t}
%+\sum_{i<j}\sum_{kk'} |c_k^{(i)}|^2|c_{k'}^{(j)}|^2\left(e^{i\left(E_{k}^{(i)}-E_{k'}^{(j)}\right)t}+e^{i\left(E_{k'}^{(j)}-E_{k}^{(i)}\right)t}\right)
%$$
%$$
%= \sum_{i} SP^{(i)}(t)+\sum_{i<j} 2\sum_{kk'} |c_k^{(i)}|^2|c_{k'}^{(j)}|^2\cos\left[\left(E_{k'}^{(j)}-E_{k}^{(i)}\right)t\right]
%$$
%$$
%=\sum_{i} SP^{(i)}(t)+\sum_{i<j} SP_I^{(ij)}(t).
%$$
%In order t
To derive an analytical expression for the interference terms $SP_I^{(ij)}$ given in (\ref{spiap}) we write
%we use the same strategy used for the one sequence case, namely we consider separately the terms for a given index distance $p$. Then, the interference contribution to the $SP$ from the $i$-th and $j$-th sub-sequences reads
\begin{align}
SP_I^{(ij)}(t)%=2\displaystyle\sum_{k,k'} |c_k^{(i)}|^2 |c_{k'}^{(j)}|^2 \cos\left[ (E_{k'}^{(j)}-E_k^{(i)})t\right]\nonumber\\
%=\displaystyle\sum_{p\in\mathbb{Z}} 2\sum_{k}  |c_k^{(i)}|^2 |c_{k+p}^{(j)}|^2 \cos\left[ (E_{k+p}^{(j)}-E_k^{(i)})t\right]
\equiv \displaystyle\sum_{p\in\mathbb{Z}} SP_{Ip}^{(ij)}(t)
=  2 \displaystyle\sum_{p\in\mathbb{Z}} \sum_{k}  g_{k}^{(i)} g_{k+p}^{(j)} \cos\left[\Omega_k^{(p)}t\right],
\label{eq:app90}
\end{align}
%Now, we use 
with $\Omega_k^{(p)}=(E_{k+p}^{(i)}-E_k^{(i)}+\delta E_{ij})$, under the assumption 
$
E_{k}^{(j)}=E_{k}^{(i)}+\delta E_{ij}. 
$
%and the observation that the components are described by a Gaussian, $ |c_k^{(i)}|^2\approx g_{k}^{(i)}=A_i \exp\left[-\frac{(E_{k}^{(i)}-\bar{E}_i)^2}{2 \sigma_i^2}\right]$, to write
%$$
%SP_{Ip}^{(ij)}(t)=2 \sum_{k}  g_{k}^{(i)} g_{k+p}^{(j)} 
%\cos\left[\Omega_k^{(p)}t\right]
%\cos\left[(E_{k+p}^{(i)}-E_k^{(i)}+\delta E_{ij})t\right]
%$$
%\begin{equation}
%SP_{Ip}^{(ij)}(t) =2 A_i A_j\sum_{k} \exp\left[-\frac{(E_{k}^{(i)}-\bar{E}_i)^2}{2 \sigma_i^2}\right]
 %\exp\left[-\frac{(E_{k+p}^{(i)}+\delta E_{ij}-\bar{E}_j)^2}{2 \sigma_j^2}\right] \cos\left[\Omega_k^{(p)}t\right],
%\label{eq:app91}
%\end{equation}
% Just like  the case of one sequence, %it can be shown 
%we  show
%that
 
 The product of the Gaussian functions $ g_{k}^{(i)} g_{k+p}^{(j)} $ leads to another single Gaussian for the frequencies $\Omega_k^{(p)}$. To demonstrate this, we express $E_k^{(i)}$ as a function of $\Omega_k^{(p)}$ employing (\ref{principalspectrum})
$$
E_k^{(i)}=\frac{(\Omega_k^{(p)} -\delta E_{ij})^2}{4 e_2 p^2}-\frac{\Omega_k^{(p)} -\delta E_{ij}+2 e_o}{2}-\frac{e_1^2 -p^2e_2^2}{4e_2}.
$$
Using this and (\ref{components}),  we obtain
%\begin{equation}
%g_{k}^{(i)} g_{k+p}^{(j)}= A_i A_j \exp\left[-\frac{(\bar{E}_i-\bar{E}_j+\Omega_k^{(p)})^2}{2(\sigma_i^2+\sigma_j^2)}\right]
%\times
%\label{eq:app92}
%\end{equation}
%$$
%\exp\left[- \frac{\sigma_i^2+\sigma_j^2}{2\sigma_i^2\sigma_j^2}\left( E_k^{(i)}-\frac{\bar{E_i} \sigma_j^2+\sigma_i^2(\bar{E}_j-\Omega_k^{(p)})}{\sigma_i^2+\sigma_j^2}\right)^2\right].
%$$
%Next, we express $E_k^{(i)}$ as a function of $\Omega_k^{(p)}$ by using the semi-classical formula $E_k^{(i)}=e_o+e_1 k+ e_2 k^2$, similarly to what  we did in \ref{App5}. The result is
%By substituting this result in the argument of the second exponential in %Eq.
%(\ref{eq:app92}), we obtain
$$
g_{k}^{(i)} g_{k+p}^{(j)}= A_i A_j \exp\left[-\frac{(\bar{E}_i-\bar{E}_j+\Omega_k^{(p)})^2}{2(\sigma_i^2+\sigma_j^2)}\right]
\exp\left[- \frac{\sigma_i^2+\sigma_j^2}{2\sigma_i^2\sigma_j^2}G\left(\Omega_k^{(p)}\right)^2\right],
$$
where $G$ %\left(\Omega_k^{(p)}\right)$
 is  a quadratic function, % $
%\left(\Omega_k^{(p)}\right)
$G=A_p \left(\Omega_k^{(p)}\right)^2+ B_p \Omega_k^{(p)}+ C_p $, with
$$
A_p=\frac{1}{4 e_2 p^2}, \ \ \ \ \ 
B_p=\frac{\sigma_i^2-\sigma_{j}^{2}}{2(\sigma_i^2+\sigma_j^2)}-\frac{\delta E_{ij}}{2 e_2 p^2},
$$
and
$$
C_p=\frac{1}{4}
\left(-\frac{e_1^2}{e_2}+ 4 e_o+e_2 p^2-\frac{4(\bar{E}_j \sigma_i^2+\bar{E}_i\sigma_j^2)}{\sigma_i^2+\sigma_j^2}
 +2 \delta E_{ij}+\frac{\delta E_{ij}^2}{e_2 p^2}\right).
$$
As in \ref{App5}, since the function $G^2$
%\left(\Omega_k^{(p)}\right)^2$
 is inside the argument of the exponential, we consider a Taylor expansion around its minimum ($\Omega_p^{ij}$) up to quadratic terms
%$$
%G\left(\Omega_k^{(p)}\right)^2\approx G\left(\Omega_p^{ij}\right)^2+ \frac{(G^2)''|_{\Omega_p^{ij}}}{2}(\Omega_k^{(p)}-\Omega_p^{ij})^2.
%$$ 
%The  minimum is given by one of the  roots  $G\left(\Omega_p^{ij}\right)=0$, because  $\left(G\left(\Omega_k^{(p)}\right)^2\right)'=2 G\left(\Omega_k^{(p)}\right) G'\left(\Omega_k^{(p)}\right)$. Then
%\begin{equation}
%\Omega_p^{ij}=\frac{-B_p\pm\sqrt{B_p^2-4 A_p C_p}}{2 A_p}, 
%\label{Omin}
%\end{equation}
%and the second derivative evaluated at the minimum is
%$$
%(G^2)''|_{\Omega_p^{ij}}=2 G\left(\Omega_p^{ij}\right)G''\left(\Omega_p^{ij}\right)+ 2 G'\left(\Omega_p^{ij}\right)^2=
%2 (2 A_p \Omega_p^{ij}+B_p)^2=2(B_p^2-4 A_pC_p).
%$$ 
%Thus, the  Taylor expansion of $G\left(\Omega_k^{(p)}\right)^2$ up to quadratic terms reads
$$
G\left(\Omega_k^{(p)}\right)^2\approx (B_p^2-4 A_pC_p)(\Omega_k^{(p)}-\Omega_p^{ij})^2 {\hbox{\ \ \ \ \ with \ \ \ \ }}  \Omega_p^{ij}=\frac{-B_p\pm\sqrt{B_p^2-4 A_p C_p}}{2 A_p}.$$
At  leading order in $e_2$
$$
\Omega_p^{ij}\approx \delta E_{ij}+p\,\omega_{ij}
{\hbox{\ \ \  \ \ and \ \ \ \ \ }}
(B_p^2-4 A_pC_p)\approx \frac{\omega_{ij}^2}{4 e_2^2 p^2},
$$
where $\omega_{ij}$ is given by
$$
\omega_{ij}=\sqrt{e_{1}^2+4e_{2}\left[\frac{\bar{E}_{j}\sigma_{i}^{2}+\bar{E}_{i}^{2}\sigma_{j}^{2}-\delta E_{ij}\sigma_{i}^{2}}{\left(\sigma_{i}^{2}+\sigma_{j}^{2}\right)}-e_{0}\right]}.
$$
The parameter $\omega_{ij}$ can be estimated  from the numerical spectrum as described below.
%In terms of the parameter $\omega_{ij}$, the product of Gaussians simplifies to
%$$
%g_{k}^{(i)} g_{k+p}^{(j)}\approx
%A_i A_j \exp\left[-\frac{(\bar{E}_i-\bar{E}_j+\Omega_k^{(p)})^2}{2(\sigma_i^2+\sigma_j^2)}\right]
%\exp\left[- \frac{\left(\Omega_k^{(p)}-(\delta E_{ij}+ p\omega_{ij})\right)^2}{2 (p\sigma_{ij})^2 }\right]
%$$

With 
$$
\sigma_{ij}=\frac{2 |e_2|\,\sigma_i \sigma_j}{\omega_{ij}\sqrt{\sigma_i^2+\sigma_j^2}}, 
$$
%As in the case of one single sequence, 
%We obtain a product of two Gaussians , with the second one being much narrower than the first. Therefore, the second gives the main contribution, while the effect of the first one is accounted for by evaluating it at 
%$\Omega_k^{(p)}= \delta E_{ij}+p \omega_{ij}$. Thus, we obtain the desired approximation
%$$
%g_{k}^{(i)} g_{k+p}^{(j)}\approx A_i A_j \exp\left[-\frac{(\bar{E}_i-\bar{E}_j+\delta E_{ij}+p\omega_{ij})^2}{2(\sigma_i^2+\sigma_j^2)}\right]
%\exp\left[- \frac{\left(\Omega_k^{(p)}-(\delta E_{ij}+ p\omega_{ij})\right)^2}{2 (p\sigma_{ij})^2 }\right].
%$$ 
%Eq.~
the terms in the sum(\ref{eq:app90}) now become
$$
 SP_{Ip}^{(ij)}(t)\approx 2 A_i A_j e^{\left[-\frac{(\bar{E}_i-\bar{E}_j+\delta E_{ij}+p\omega_{ij})^2}{2(\sigma_i^2+\sigma_j^2)}\right]}
\sum_{k}\exp\left[- \frac{\left(\Omega_k^{(p)}-(\delta E_{ij}+ p\omega_{ij})\right)^2}{2 (p\sigma_{ij})^2 }\right] \cos\left[\Omega_k^{(p)}t\right].
$$
From  (\ref{principalspectrum})%$E_{k}^{(i)}=e_o+e_1 k+e_2 k^2$
, we obtain $\Delta \Omega^{(p)}=\Omega_{k+1}^{(p)}-\Omega_{k}^{(p)}=2 p e_2$,  
%the sum in the above expression  can be approximated by an integral as 
to approximate  the previous
 sum  by an integral that can be calculated as in \ref{App7},
%$$
%\sum_{k}\exp\left[- \frac{\left(\Omega_k^{(p)}-(\delta E_{ij}+ p\omega_{ij})\right)^2}{2 (p\sigma_{ij})^2 }\right] \cos\left[\Omega_k^{(p)}t\right]
%$$
%$$
%\approx\frac{1}{2p |e_2|} \int \exp\left[- \frac{\left(\Omega-(\delta E_{ij}+ p\omega_{ij})\right)^2}{2(p\sigma_{ij})^2 }\right] \cos\left[\Omega t\right] d \Omega.
%$$
%The integral can be calculated straightforwardly as we did in \ref{App7}, resulting in
%$$
%\frac{1}{2 p |e_2|} \sqrt{2 \pi} \sigma_{ij} \exp\left[-\frac{(\sigma_{ij}t)^2}{2}\right] \cos[(\delta E_{ij}+ p\omega_{ij}) t].
%$$
%By substituting this result and the expression for $\sigma_{ij}$, we obtain
$$
SP_{Ip}^{(ij)}(t)\!\approx\!
 \frac{2 A_i A_j\sqrt{2\pi}\sigma_i\sigma_j }{\omega_{ij}\sqrt{\sigma_i^2+\sigma_j^2}} e^{\left[-\frac{(p\sigma_{ij}t)^2}{2}\right]} 
\exp\!\!\left[-\frac{(\!\bar{E}_i-\!\bar{E}_j+\!\delta E_{ij}+\!p\omega_{ij})^2}{2(\sigma_i^2+\sigma_j^2)}\right]  \cos[(\delta E_{ij}+ p\omega_{ij}) t].
$$
%We now  substitute this  result  in  %Eq.~
%(\ref{eq:app90}) to obtain the final result for the interference term between the $i$-th and $j$-th sub-sequences
%$$
%SP_{I}^{(ij)}(t)=
%$$
%$$ \frac{2 A_i A_j\sqrt{2\pi}\sigma_i\sigma_j}{\omega_{ij}\sqrt{\sigma_i^2+\sigma_j^2}}\sum_{p\in\mathbb{Z}} \exp\left[-\frac{(p\sigma_{ij}t)^2}{2}\right] 
% \exp\left[-\frac{(\bar{E}_i-\bar{E}_j+\delta E_{ij}+p\omega_{ij})^2}{2(\sigma_i^2+\sigma_j^2)}\right] \cos[(\delta E_{ij}+ p\omega_{ij}) t].
%$$

Finally, we present a simple way to estimate the parameter $\omega_{ij}$.  According to the  discussion above, the maximum of the product $g_{k}^{(i)} g_{k+1}^{(j)}$  is $\Omega_{p=1}^{(ij)}=\delta E_{ij} + \omega_{ij}$. On the other hand,  the product of two Gaussian functions, considering  the energy as  a continuous variable $x$ and a fixed parameter $\Delta$,   $g^{(i)}(x) g^{(j)}(x+\Delta)= A_i A_j\exp\left[-\frac{(x-\bar{E_i})^2}{2 \sigma_i^2}\right] 
 \exp\left[-\frac{(x+\Delta-\bar{E_j})^2}{2 \sigma_j^2}\right],$ 
acquires its maximum value at 
$
x_{max}=E_{ij}^{(I)}-\frac{\sigma_i^2}{\sigma_i^2+\sigma_j^2}\Delta, 
$
with
\begin{equation}
E_{ij}^{(I)}=\frac{\bar{E}_i \sigma_j^2+\bar{E}_j \sigma_i^2}{\sigma_i^2+\sigma_j^2}.
\label{eq:app93}
\end{equation}
For the argument of the second Gaussian, we sum  %$\Delta$ to  $x_{max}$,
$x_{max}+\Delta= E_{ij}^{(I)}+\frac{\sigma_j^2}{\sigma_i^2+\sigma_j^2}\Delta$.
Therefore the pair of continuous energies maximizing the product of Gaussians are located to the right and to the left of $E_{ij}^{(I)}$. Returning to the  discrete spectrum, the best  approximation to this pair of energies is given by the pair of consecutive discrete energies satisfying 
\begin{equation}
E_{k_I}^{(i)}\leq E_{ij}^{(I)}\leq E_{k_I+1}^{(i)}. 
\label{eq:app94}
\end{equation}
 From this simple observation,  the frequency maximizing the Gaussian product can be approximated by $\Omega_{p=1}^{(ij)}\approx E_{k_I+1}^{(i)}+\delta E_{ij}-E_{k_I}^{(i)}$. Comparing  this with the equivalent expression in terms of $\omega_{ij}$, that is $\Omega_{p=1}^{(ij)}=\delta E_{ij} + \omega_{ij}$, allows for the  estimation 
$$
\omega_{ij}\approx E_{k_I+1}^{(i)}-E_{k_I}^{(i)}.
$$
This provides a %simple
 way to estimate $\omega_{ij}$ from the numerical spectrum by using %Eq.~
(\ref{eq:app93}) and %the condition 
(\ref{eq:app94}).

%%%%%%%%%%%%%%%%%%%%%%%%%%%

%%%%%%%%%%%%%%%%%%%%%%%%%%%%%%%%%%%%%%%%%%%%%%%%%%%
\section*{References}
\bibliographystyle{iopart-num}

%\bibliography{CSsp}

\begin{thebibliography}{10}
\expandafter\ifx\csname url\endcsname\relax
  \def\url#1{{\tt #1}}\fi
\expandafter\ifx\csname urlprefix\endcsname\relax\def\urlprefix{URL }\fi
\providecommand{\eprint}[2][]{\url{#2}}
% Bibliography created with iopart-num v2.1
% /biblio/bibtex/contrib/iopart-num



\bibitem{Chu2002}
Chu S 2002 {\em Nature (London)\/} {\bf 416} 206
 % \urlprefix\url{http://dx.doi.org/10.1038/416206a}

\bibitem{BernienARXIV}
Bernien H, Schwartz S, Keesling A, Levine H, Omran A, Pichler H, Choi S, Zibrov
  A~S, Endres M, Greiner M, Vuleti\'c V and Lukin M~D Probing many-body
  dynamics on a 51-atom quantum simulator arXiv:1707.04344

\bibitem{Blatt2012}
Blatt R and Roos C~F 2012 {\em Nat. Phys.\/} {\bf 8} 277--284

\bibitem{Richerme2014}
Richerme P, Gong Z~X, Lee A, Senko C, Smith J, Foss-Feig M, Michalakis S,
  Gorshkov A~V and Monroe C 2014 {\em Nature (London)\/} {\bf 511} 198--201

\bibitem{WeiARXIV}
Wei K~X, Ramanathan C and Cappellaro P 2018 {\em Phys. Rev. Lett.} {\bf 120} 070501 %Exploring localization in nuclear spin
  chains arXiv:1612.05249

\bibitem{Lipkin1965a}
Lipkin H~J, Meshkov N and Glick A~J 1965 {\em Nucl. Phys.\/} {\bf 62} 188--198

\bibitem{Lipkin1965b}
Meshkov N, Glick A~J and Lipkin H~J 1965 {\em Nucl. Phys.\/} {\bf 62} 199--210

\bibitem{Lipkin1965c}
Glick A~J, Lipkin H~J and Meshkov N 1965 {\em Nucl. Phys.\/} {\bf 62} 211--224

\bibitem{Dicke54}
Dicke R~H 1954 {\em Phys. Rev.\/} {\bf 93} 99
  %\urlprefix\url{https://link.aps.org/doi/10.1103/PhysRev.93.99}

\bibitem{Hepp73}
Hepp K and Lieb E~H 1973 {\em Ann. Phys. (N.Y.)\/} {\bf 76} 360% -- 404 ISSN
  0003-4916
  %\urlprefix\url{http://www.sciencedirect.com/science/article/pii/0003491673900390}

\bibitem{Wang1973}
Wang Y~K and Hioe F~T 1973 {\em Phys. Rev. A\/} {\bf 7}(3) 831--836
  %\urlprefix\url{https://link.aps.org/doi/10.1103/PhysRevA.7.831}

\bibitem{Milburn1997} G. J. Milburn, J. Corney, E. M. Wright, and D. F. Walls 1997
{\em Phys. Rev. A} {\bf 55}  4318.

\bibitem{Steel1998} M. J. Steel and M. J. Collett 1998
{\em Phys. Rev. A} {\bf 57} 2920.

\bibitem{Zibold2010}
Zibold T, Nicklas E, Gross C and Oberthaler M~K 2010 {\em Phys. Rev. Lett.\/}
  {\bf 105}(20) 204101

\bibitem{Baumann2010}
Baumann K, Guerlin C, Brennecke F and Esslinger T 2010 {\em Nature (London)\/}
  {\bf 464} 1301

\bibitem{Baumann2011}
Baumann K, Mottl R, Brennecke F and Esslinger T 2011 {\em Phys. Rev. Lett.\/}
  {\bf 107} 140402
  %\urlprefix\url{https://link.aps.org/doi/10.1103/PhysRevLett.107.140402}
\bibitem{Campbell16}  Campbell S 2016 	{\em Phys. Rev. B} {\bf  94} 184403 

\bibitem{Wang17}  Wang Q,  Quan H~T 2017 {\em Phys. Rev. E} {\bf 96} 032142

\bibitem{SantosBernal2015}
Santos L~F and P\'erez-Bernal F 2015 {\em Phys. Rev. A\/} {\bf 92} 050101
  %\urlprefix\url{http://link.aps.org/doi/10.1103/PhysRevA.92.050101}

\bibitem{Santos2016}
Santos L~F, T\'avora M and P\'erez-Bernal F 2016 {\em Phys. Rev. A\/} {\bf
  94} 012113
  %\urlprefix\url{https://link.aps.org/doi/10.1103/PhysRevA.94.012113}

\bibitem{Fernandez2011}
P\'erez-Fern\'andez P, Cejnar P, Arias J~M, Dukelsky J, Garc\'{i}a-Ramos J~E
  and Rela\~no A 2011 {\em Phys. Rev. A\/} {\bf 83}(3) 033802
  
 

\bibitem{DQPT} %Dynamical Quantum Phase Transitions in the Transverse-Field Ising Model 
Heyl M,  Polkovnikov A, and Kehrein S 2013 {\em  Phys. Rev. Lett. } {\bf 110}  135704

\bibitem{DQPT2}   Heyl M 2018 {\em Rep. Prog. Phys.} {\bf 81}  054001

\bibitem{Alhassid92}  Alhassid Y and  Levine R~D 1992  {\em Phys. Rev. A}  {\bf 46} 4650 
\bibitem{Torres2017b}  E. J. Torres-Herrera and L. F. Santos, Phil. Trans. R.
Soc. A 375, 20160434 (2017). 

\bibitem{Fonda78} L. Fonda, G. C. Ghirardi, and A. Rimini, Rep. Prog. Phys., 41, 587 (1978).


\bibitem{Ketz92} R. Ketzmerick, G. Petschel, and T. Geisel, Phys. Rev. Lett. 69,
695 (1992).

\bibitem{Khalfin58} L. A. Khalfin, Sov. Phys. JETP 6, 1053 (1958);
J. G. Muga, A. Ruschhaupt, and A. del Campo, Time in Quantum Mechanics, vol. 2 (Springer, London 2009)

\bibitem{Peres1984}  Peres A 1984 {\em  Phys. Rev. A} {\bf 30} 161 

\bibitem{Gorin2006}  Gorin T,  Prosen T,  Seligman T~H,  \u{Z}nidari\u{c} M 2006 {\em Phys. Rep.} {\bf 435} 33
  
\bibitem{Prosen2002} Prosen T and  \u{Z}nidari\u{c} M 2002 {\em J. Phys. A: Math. Gen.} {\bf 35}  1455
  
\bibitem{Prosen2003} Prosen T,  Seligman T~H,  \u{Z}nidari\u{c} M 2003  	{\em Phys. Rev. A} {\bf 67} 042112  

%%%%%%************AGRUPAR
\bibitem{Torres2014PRA}
Torres-Herrera E~J and Santos L~F 2014 {\em Phys. Rev. A\/} {\bf 89} 043620, 
\bibitem{Torres2014NJP}
Torres-Herrera E~J, Vyas M and Santos L~F 2014 {\em New J. Phys.\/} {\bf 16}
  063010, 
  %\urlprefix\url{http://iopscience.iop.org/article/10.1088/1367-2630/16/6/063010/meta#}
\bibitem{Torres2014PRE}
Torres-Herrera E~J and Santos L~F 2014 {\em Phys. Rev. E\/} {\bf 89} 062110

\bibitem{Torres2014PRAb}
Torres-Herrera E~J and Santos L~F 2014 {\em Phys. Rev. A\/} {\bf 90} 033623
%%%%%%%%%%%%%%%%%%%%%%%%% 

\bibitem{Flamabum2001}  Flambaum V~V and  Izrailev F~M 2001  {\em Phys. Rev. E} {\bf 64} 026124 

\bibitem{Izrailev2006}  Izrailev F~M and  Casta\~neda-Mendoza A 2006 {\em Phys. Lett. A} {\bf 350} 355



%%%%%%%%%%%%%%AGRUPAR***********
%\bibitem{Torres2015}
%Torres-Herrera E~J and Santos L~F 2015 {\em Phys. Rev. B\/} {\bf 92} 014208
  %\urlprefix\url{http://link.aps.org/doi/10.1103/PhysRevB.92.014208}

\bibitem{Tavora2016}
T\'avora M, Torres-Herrera E~J and Santos L~F 2016 {\em Phys. Rev. A\/} {\bf
  94} 041603,
  %\urlprefix\url{http://link.aps.org/doi/10.1103/PhysRevA.94.041603}
\bibitem{Tavora2017}
T\'avora M, Torres-Herrera E~J and Santos L~F 2017 {\em Phys. Rev. A\/} {\bf
  95} 013604
  %\urlprefix\url{http://link.aps.org/doi/10.1103/PhysRevA.95.013604}

\bibitem{Torres2017}
Torres-Herrera E~J and Santos L~F 2017 {\em Ann. Phys. (Berl.)\/} {\bf 529}
  1600284 %ISSN 1521-3889
  %\urlprefix\url{http://dx.doi.org/10.1002/andp.201600284}
%%%%%%%***********************************************

\bibitem{KlauderBook}
Klauder J~R and Skagerstam B-S 1985 {\em Coherent States: Applications in Physics
  and Mathematical Physics\/} (Singapore: World Scientific)
  
\bibitem{Gilmore90} %Coherent states: Theory and some applications
Zhang W-M,  Feng D~H, and  Gilmore R 1990 {\em Rev. Mod. Phys.} {\bf 62} 867 

\bibitem{Altland2012NJP}
Altland A and Haake F 2012 {\em New J. Phys.\/} {\bf 14} 073011
  %\urlprefix\url{http://stacks.iop.org/1367-2630/14/i=7/a=073011}

\bibitem{Altland2012PRL}
Altland A and Haake F 2012 {\em Phys. Rev. Lett.\/} {\bf 108} 073601
  %\urlprefix\url{https://link.aps.org/doi/10.1103/PhysRevLett.108.073601}
  
\bibitem{Bakemeier13}  Bakemeier L,  Alvermann A and  Fehske H 2013 {\em Phys. Rev. A} {\bf 88} 043835 



\bibitem{Leichtle1996}
Leichtle C, Averbukh I~S and Schleich W~P 1996 {\em Phys. Rev. A\/} {\bf 54}
  5299--5312 %\urlprefix\url{https://link.aps.org/doi/10.1103/PhysRevA.54.5299}

\bibitem{Kloc18}  Kloc M,  Str\'ansk\'y P,  Cejnar P 2018  {\em	Phys. Rev. A 98} {\bf 013836}

\bibitem{HUMBERTO}V\'azquez-S\'anchez H, Lerma-Hern\'andez S, work in progress

\bibitem{Relano16}
Rela\~no A, Bastarrachea-Magnani M~A and Lerma-Hern\'andez S 2016 {\em EPL\/}
  {\bf 116} 50005
  %\urlprefix\url{http://stacks.iop.org/0295-5075/116/i=5/a=50005}

\bibitem{BastaJPA17}
Bastarrachea-Magnani M~A, Rela\~no A, Lerma-Hern\'andez S, L\'opez-del{-}Carpio
  B, Ch\'avez-Carlos J and Hirsch J~G 2016 {\em J. Phys. A\/} {\bf 50}

\bibitem{Castanos2005}
Casta\~nos O, L\'opez-Pe\~na R, Hirsch J~G and L\'opez-Moreno E 2005 {\em Phys.
  Rev. B\/} {\bf 72} 012406
  %\urlprefix\url{https://link.aps.org/doi/10.1103/PhysRevB.72.012406}

\bibitem{Emary2003PRL}
Emary C and Brandes T 2003 {\em Phys. Rev. Lett.\/} {\bf 90} 044101
  %\urlprefix\url{https://link.aps.org/doi/10.1103/PhysRevLett.90.044101}

\bibitem{Emary2003}
Emary C and Brandes T 2003 {\em Phys. Rev. E\/} {\bf 67} 066203
  %\urlprefix\url{https://link.aps.org/doi/10.1103/PhysRevE.67.066203}
  
\bibitem{Ribeiro2006}
Ribeiro A~D, de~Aguiar M~A~M and de~Toledo~Piza A~F~R 2006 {\em J. Phys. A\/}
  {\bf 39} 3085 %\urlprefix\url{http://stacks.iop.org/0305-4470/39/i=12/a=016}

\bibitem{Larson17}
Larson J and Irish E~K 2017 {\em J. Phys. A\/} {\bf 50} 174002
  %\urlprefix\url{http://stacks.iop.org/1751-8121/50/i=17/a=174002}  

\bibitem{CastanosProceed}
Casta\~{n}os O, Nahmad-Achar E, L\'{o}pez-Pe\~{n}a R and Hirsch J~G 2010 {\em
  AIP Conf. Proc.\/} {\bf 1323} 40--51
  %\urlprefix\url{http://aip.scitation.org/doi/abs/10.1063/1.3537864}

\bibitem{Caprio2008} Caprio M~A,  Cejnar P,  Iachello F (2008) {\em Ann. Phys. (N.Y.)} {\bf 323}  1106

\bibitem{Ribeiro2008}
Ribeiro P, Vidal J and Mosseri R 2008 {\em Phys. Rev. E\/} {\bf 78} 021106

\bibitem{Bastarrachea14PRA-1}
Bastarrachea-Magnani M~A, Lerma-Hern\'andez S and Hirsch J~G 2014 {\em Phys.
  Rev. A\/} {\bf 89} 032101
  %\urlprefix\url{https://link.aps.org/doi/10.1103/PhysRevA.89.032101}


\bibitem{Stransky2014}
Str\'ansk\'y P, Macek M and Cejnar P 2014 {\em Ann. Phys. (N.Y.)\/} {\bf 345}
  73--97

\bibitem{Stransky2015}
Str\'ansk\'y P, Macek M, Leviatan A and Cejnar P 2015 {\em Ann. Phys. (N.Y.)\/}
  {\bf 356} 57 %-- 82 ISSN 0003-4916
  %\urlprefix\url{http://www.sciencedirect.com/science/article/pii/S0003491615000767}

\bibitem{Bernal2017}
P\'erez-Bernal F and Santos L~F 2017 {\em Fortschr. Phys.\/} {\bf 65} 1600035

\bibitem{Engelhardt2015}
Engelhardt G, Bastidas V~M, Kopylov W and Brandes T 2015 {\em Phys. Rev. A\/}
  {\bf 91} 013631
  
%\bibitem{Puebla2013}
%Puebla R, Rela\~no A and Retamosa J 2013 {\em Phys. Rev. A\/} {\bf 87}(2)
%  023819 %\urlprefix\url{https://link.aps.org/doi/10.1103/PhysRevA.87.023819} 
  


\bibitem{Chen2008}
Chen Q~H, Zhang Y~Y, Liu T and Wang K~L 2008 {\em Phys. Rev. A\/} {\bf 78}
  051801 %\urlprefix\url{https://link.aps.org/doi/10.1103/PhysRevA.78.051801}

\bibitem{Liu2009}
Liu T, Zhang Y~Y, Chen Q~H and Wang K~L 2009 {\em Phys. Rev. A\/} {\bf 80}
  023810 %\urlprefix\url{https://link.aps.org/doi/10.1103/PhysRevA.80.023810}

\bibitem{Bastarrachea2014PSa}
Bastarrachea-Magnani M~A and Hirsch J~G 2014 {\em Phys. Scr.\/} {\bf 2014}
  014005 %\urlprefix\url{http://stacks.iop.org/1402-4896/2014/i=T160/a=014005}

\bibitem{Bastarrachea2014PSb}
Hirsch J~G and Bastarrachea-Magnani M~A 2014 {\em Phys. Scr.\/} {\bf 2014}
  014018 %\urlprefix\url{http://stacks.iop.org/1402-4896/2014/i=T160/a=014018}  

\bibitem{Bastarrachea2016a}
Bastarrachea-Magnani M~A, L\'opez-del{-}Carpio B, Ch\'avez-Carlos J,
  Lerma-Hern\'andez S and Hirsch J~G 2016 {\em Phys. Rev. E\/} {\bf 93}
  022215 %\urlprefix\url{http://link.aps.org/doi/10.1103/PhysRevE.93.022215}



\bibitem{TorresKollmar2015}
Torres-Herrera E~J, Kollmar D and Santos L~F 2015 {\em Phys. Scr. T\/} {\bf
  165} 014018




\bibitem{Schliemann2015}
Schliemann J 2015 {\em Phys. Rev. A\/} {\bf 92} 022108
  %\urlprefix\url{https://link.aps.org/doi/10.1103/PhysRevA.92.022108}

\bibitem{BastarracheaPS2017}
Bastarrachea-Magnani M~A, L\'opez-del{-}Carpio B, Ch\'avez-Carlos J,
  Lerma-Hern\'andez S and Hirsch J~G 2017 {\em Phys. Scr.\/} {\bf 92} 054003
  %\urlprefix\url{http://stacks.iop.org/1402-4896/92/i=5/a=054003}



\bibitem{Jacobi}
Weisstein E~W {\it Jacobi theta functions} MathWorld-A Wolfram Web Resource.
  http://mathworld.wolfram.com/JacobiThetaFunctions.html [September 2017]

\bibitem{Bastarrachea14PRA-2}
Bastarrachea-Magnani M~A, Lerma-Hern\'andez S and Hirsch J~G 2014 {\em Phys.
  Rev. A\/} {\bf 89} 032102
  %\urlprefix\url{https://link.aps.org/doi/10.1103/PhysRevA.89.032102}


\bibitem{Chavez2016}
Ch\'avez-Carlos J, Bastarrachea-Magnani M~A, Lerma-Hern\'andez S and Hirsch J~G
  2016 {\em Phys. Rev. E\/} {\bf 94} 022209
  %\urlprefix\url{https://link.aps.org/doi/10.1103/PhysRevE.94.022209}

\bibitem{LermaAIP18} Lerma-Hern\'andez S, Ch\'avez-Carlos J, Bastarrachea-Magnani M A, L\'opez-del-Carpio B and Hirsch J G 2018,  {\em AIP Conf. Proc.} {\bf 1950} 030002


\bibitem{ReichlBook}
Reichl L~E 2004 {\em The transition to chaos: conservative classical systems
  and quantum manifestations\/} (New York: Springer)


\bibitem{Wisniacki2015}
Wisniacki D~A and Schlagheck P 2015 {\em Phys. Rev. E\/} {\bf 92} 062923
  %\urlprefix\url{https://link.aps.org/doi/10.1103/PhysRevE.92.062923}

\bibitem {Jaako16} Jaako T, Xiang Z-L, Garcia-Ripoll J~J, and  Rabl P 2016 {\em Phys. Rev. A} {\bf 94} 033850 
\bibitem{Felicetti15}  Felicetti S,  Pedernales, J~S,  Egusquiza I~L,  Romero G,  Lamata L,  Braak D, and  Solano E 2015 {\em Phys. Rev. A} {\bf 92} 033817 

\bibitem{Penna17} Penna V,  Raffa F~A and  Franzosi R 2018 {\em J. Phys. A: Math. Theor.} {\bf 51}  045301 


%%\bibitem{BerryProceed}
%Berry M~V 1991 Some quantum-to-classical asymptotics {\em Chaos and Quantum
%  Physics\/} Les Houches Lecture Series LII, 1989 ed Giannoni M~J, Voros A and
%  Zinn-Justin J (Amsterdam: Elsevier Science Publishers)



%%\bibitem{Marklof1999}
%%Marklof J 1999 Theta sums, eisenstein series, and the semiclassical dynamics of
%%  a precessing spin {\em Emerging Applications of Number Theory\/} ed Hejhal
%%  D~A, Friedman J, Gutzwiller M~C and Odlyzko A~M (New York, NY: Springer New
%% York) pp 405--450

%%\bibitem{Yuan2012}
%%Yuan Z~G, Zhang P, Li S~S, Jing J and Kong L~B 2012 {\em Phys. Rev. A\/} {\bf
%%  85}(4) 044102


%\bibitem{Cejnar2015}
%Cejnar P, Str\'ansk\'y P and Kloc M 2015 {\em Phys. Scr.\/} {\bf 90} 114015
  %\urlprefix\url{http://stacks.iop.org/1402-4896/90/i=11/a=114015}



\end{thebibliography}

%%%%%%%%%%%%%%%%%%%%%%%%%%%%%%%%%%%%%%%%%%%%%%%%%%%%%%%%%%%%%%%%%%%%%%%%
%%%%%%%%%%%%%%%%%%%%%%%%%%%%%%%%%%%%%%%%%%%%%%%%%%%%%%%%%%%%%%%%%%%%%%%%%

\end{document}